\documentclass[12pt,a4paper,twoside]{article}

\usepackage[applemac]{inputenc}
\usepackage[english]{babel}
\usepackage[T1]{fontenc}
\usepackage{lmodern}
\usepackage{geometry} 
\usepackage[intlimits,sumlimits]{amsmath}
\usepackage{amssymb}
\usepackage[amsthm,thmmarks]{ntheorem}
\usepackage{mathrsfs}
\usepackage{bm}
\usepackage{xpatch}
\usepackage{dsfont}
\usepackage{siunitx} 
\usepackage{scalerel,stackengine}
\usepackage{aligned-overset}
\usepackage{algorithm}
\usepackage{algpseudocode}
\usepackage{multicol}
\usepackage{booktabs}
\usepackage{multirow}
\usepackage{makecell}

\usepackage[backend=bibtex, style=authoryear, firstinits=true, maxcitenames=2, sorting=nty, isbn=false, dashed=false]{biblatex}
\DeclareNameAlias{sortname}{last-first}
\DeclareFieldFormat*{title}{#1}
\DeclareFieldFormat*{edition}{#1 edition}
\DeclareDelimFormat[cite,parencite]{nameyeardelim}{\textcolor{black}{\addcomma\space}}

\DeclareFieldFormat{citehyperref}{%
   \DeclareFieldAlias{bibhyperref}{noformat}
   \bibhyperref{#1}}

\savebibmacro{cite}

\renewbibmacro*{cite}{%
   \printtext[citehyperref]{%
      \restorebibmacro{cite}%
      \usebibmacro{cite}}}
\usepackage{xpatch}
\xpatchbibmacro{textcite}
  {\printnames{labelname}}
  {\printtext[bibhyperref]{\printnames{labelname}}}
  {}{}

\usepackage{caption}
\usepackage{subcaption}
\usepackage{paralist}
\usepackage{graphicx}
\usepackage{colortbl}
\usepackage[percent]{overpic}
\usepackage{mathtools}
\usepackage[usenames,dvipsnames,svgnames,table]{xcolor}
\usepackage{pgf,tikz}
\usetikzlibrary{patterns}
\usetikzlibrary{quotes,babel,angles}
\usetikzlibrary{arrows}
\usepackage{rotating}
\usepackage{pgfplots}
\usepackage{wrapfig}
\usepackage[most]{tcolorbox}
\usepackage[compact]{titlesec}
\usepackage{fancyhdr}

\definecolor{UHwhite}{RGB}{255,255,255}
\definecolor{UHorange}{RGB}{255,128,0}
\definecolor{UHred}{RGB}{202,53,56}
\definecolor{UHgruen}{RGB}{161,176,45}
\definecolor{UHdarkblue}{RGB}{6,68,107}
\definecolor{UHblue}{RGB}{24,105,183}
\definecolor{UHmidblue}{RGB}{17,119,182}
\definecolor{UHcyan}{RGB}{87,168,211}
\definecolor{UHbrightblue}{RGB}{160,206,234}
\definecolor{UHdarkgrey}{RGB}{69,69,69}
\definecolor{UHgrey}{RGB}{148,148,148}
\definecolor{UHbrightgrey}{RGB}{215,215,215}
\definecolor{UHbeige}{RGB}{235,230,215}

\definecolor{myred}{RGB}{178,34,34}
\definecolor{mygrey}{RGB}{69,69,69}

\numberwithin{equation}{section}
\definecolor{blau}{RGB}{24,105,183}
\usepackage[colorlinks=False,
pdfpagelabels,
pdfstartview = FitH,
bookmarksopen = true,
bookmarksnumbered = true,
colorlinks   = true,
linkcolor = black,
citecolor = UHblue,
plainpages = false,
hypertexnames = false,
linkbordercolor = UHblue,
urlcolor     = UHblue,
citebordercolor=white,
urlbordercolor=white] {hyperref}
\usepackage{orcidlink}

\usepackage{sectsty}
\usepackage{booktabs}

\usepackage[justification=centering]{caption}
\usepackage{dcolumn}
\usepackage{framed}
\newcolumntype{2}{D{.}{}{3.0}}
\makeatletter 
\renewcommand*{\@pnumwidth}{3em}      
\makeatother
\makeatletter
\renewcommand{\p@envcount}{\thesubsection.}
\makeatother

\newcommand{\R}{\mathbb{R}}

\newcommand{\mng}[1]{\left \{ #1 \right \}}

\newcommand{\spn}[1]{\mathrm{span}\kern-0.4ex\left(#1\right)}
\newcommand{\supp}[1]{\mathrm{supp}\kern-0.4ex\left(#1\right)}
\newcommand{\bild}[1]{\mathrm{Bild}\kern-0.4ex\left( #1\right)}
\newcommand{\krn}[1]{\mathrm{Kern}\kern-0.4ex\left( #1\right)}
\newcommand{\rang}[1]{\mathrm{rang}\kern-0.4ex\left( #1\right)}
\newcommand{\spur}[1]{\mathrm{Spur}\kern-0.4ex\left( #1\right)}

\newcommand{\ggt}[1]{\mathrm{ggT}\kern-0.4ex\left( #1\right)}
\newcommand{\kgv}[1]{\mathrm{kgV}\kern-0.4ex\left( #1\right)}

\newcommand{\abs}[1]{\left| #1 \right|} 
\newcommand{\norm}[1]{\left\| #1 \right\|}

\newcommand{\grpGL}[1]{\mathrm{GL}\kern-0.4ex\left( #1\right)}
\newcommand{\grpSL}[1]{\mathrm{SL}\kern-0.4ex\left( #1\right)}
\newcommand{\grpO}[1]{\mathrm{O}\kern-0.4ex\left( #1\right)}
\newcommand{\grpSO}[1]{\mathrm{SO}\kern-0.4ex\left( #1\right)}
\newcommand{\grpU}[1]{\mathrm{U}\kern-0.4ex\left( #1\right)}
\newcommand{\grpSU}[1]{\mathrm{SU}\kern-0.4ex\left( #1\right)}

\renewcommand{\P}[1]{\mathbb{P}\kern-0.4ex\left( #1\right)}
\newcommand{\E}[1]{\mathbb{E}\kern-0.4ex\left( #1\right)}

\DeclareMathOperator*{\argmin}{arg\,min}

\makeatletter
\renewenvironment{abstract}{%
    \if@twocolumn
      \section*{\abstractname}%
    \else 
      \begin{center}%
        {\bfseries \abstractname\vspace{\z@}}%
      \end{center}%
      \quotation
    \fi}
    {\if@twocolumn\else\endquotation\fi}
\makeatother

\begin{document}
\sloppy


\title{\fontsize{24}{30} \selectfont \textbf{Gradient-Boosted Generalized Linear Models for Conditional Vine Copulas}}

\author{David Jobst\,\orcidlink{0000-0002-2014-3530}\thanks{Corresponding author, University of Hildesheim, Institute of Mathematics and Applied Informatics, Samelsonplatz 1, 31141 Hildesheim, Germany, \texttt{\href{mailto:jobstd@uni-hildesheim.de}{jobstd@uni-hildesheim.de}}},\and  Annette M\"oller\,\orcidlink{0000-0001-9386-1691}\thanks{Bielefeld University, Faculty of Business Administration and Economics, Universit\"atsstra{\ss}e 25, 33615 Bielefeld, Germany, \texttt{\href{mailto:annette.moeller@uni-bielefeld.de}{annette.moeller@uni-bielefeld.de}}} 
\and J\"urgen Gro{\ss}\,\orcidlink{0000-0002-3861-4708}\thanks{University of Hildesheim, Institute of Mathematics and Applied Informatics, Samelsonplatz 1, 31141 Hildesheim, Germany, \texttt{\href{mailto:juergen.gross@uni-hildesheim.de}{juergen.gross@uni-hildesheim.de}}}}
\maketitle
\thispagestyle{empty}

\begin{abstract}
\small Vine copulas are flexible dependence models using bivariate copulas as building blocks. If the parameters of the bivariate copulas in the vine copula depend on covariates, one obtains a conditional vine copula. We propose an extension for the estimation of continuous conditional vine copulas, where the parameters of continuous conditional bivariate copulas are estimated sequentially and separately via gradient-boosting. For this purpose, we link covariates via generalized linear models (GLMs) to Kendall's $\tau$ correlation coefficient from which the corresponding copula parameter can be obtained. Consequently, the gradient-boosting algorithm estimates the copula parameters providing a natural covariate selection. In a second step, an additional covariate deselection procedure is applied. The performance of the gradient-boosted conditional vine copulas is illustrated in a simulation study. Linear covariate effects in low- and high-dimensional settings are investigated for the conditional bivariate copulas separately and for conditional vine copulas. Moreover, the gradient-boosted conditional vine copulas are applied to the temporal postprocessing of ensemble weather forecasts in a low-dimensional setting. The results show, that our suggested method is able to outperform the benchmark methods and identifies temporal correlations better. Eventually, we provide an \texttt{R}-package called \texttt{boostCopula} for this method.

\end{abstract}

\newpage

\section{Introduction}
\label{sec: Introduction}

In plenty of practical applications, multivariate data needs to be modeled by multivariate statistical models, such as e.g. the multivariate normal distribution. Unfortunately, these models are sometimes too inflexible in their marginal and joint behavior, as they e.g. assume, that all marginal distributions belong to the same family. Copulas allow to overcome this issue, as they model the dependence structure between the variables independently of the univariate marginal distributions. 


Often parametric copulas are employed, where the parameters are mostly treated as a constant not related to any covariates. As a natural extension, the copula parameter can depend on covariates which yields to a so-called \textit{conditional copula} originally introduced by \textcite{Patton2002}. Later, non-parametric \parencite{Gijbels2011, Veraverbeke2011}, semi-parametric \parencite{Acar2010} and Bayesian \parencite{Craiu2012, Sabeti2014} formulations of the conditional copula have been investigated. \textcite{Vatter2015} modeled the copula parameters using generalized additive models (GAMs, \cite{Hastie1986, Hastie1990, Wood2017}). While a two-step approach called inference for margins (IFM, \cite{Joe1996a}), where first the marginal distributions and afterwards the conditional copulas are estimated is frequently applied, \textcite{Klein2015, Radice2015, Marra2017} estimate the marginal distributions and conditional copula parameters simultaneously. 

In particular, if the amount of covariates $p$ becomes large, it is not always straightforward for conditional bivariate copulas to include the most informative and exclude the non-informative covariates. 
One way to tackle these issues is model-based boosting \parencites{Buehlmann2003, Buehlmann2007} which is based on a functional gradient-boosting idea \parencite{Friedman2001}. This approach provides a natural variable selection, shrinks effect estimates leading to a better prediction accuracy, and is robust against multicollinearity issues \parencite{Mayr2018}. In a copula regression framework, \textcite{Hans2022} developed the so called \textit{boosted copula regression} for continuous bivariate responses in a biomedical context. The parameters of the marginal distributions and the parameter of the copula are estimated jointly using the gradient-boosting approach. Recently, \textcite{Sanchez2024} extended this idea for bivariate binary, discrete and mixed responses with application to biomedical data. 


However, this boosted copula regression approach comes along with some drawbacks. Most notably, it is restricted to bivariate copulas only. Extending this approach in its current fashion to multivariate copulas might be time-consuming due to the modeling of the marginal distributions and a conditional copula jointly. Furthermore, the zoo of multivariate parametric copulas is limited \parencite{Genest2009} and they usually assume homogeneous dependence among the variables, which would restrict the boosted copula regression in its flexibility for dependence modeling. Additionally, the gradient-boosting generally tends to include too many covariates in a low-dimensional setting, which can lead to a slow overfitting \parencite{Hans2022}. 
Eventually, a larger set of more flexbile copula families might be beneficial for different applications. 

To overcome these issues, we follow the two-step approach (IFM), where the marginal distributions can be individually individually specified by the user in advance. In the next step a conditional multivariate copula is estimated. Latter is implemented as follows:
\begin{compactenum}[(i)]
    \item We suggest the use of a \textit{vine copula} as a flexible copula class. A vine copula can be obtained by decomposing a copula into a cascade of bivariate copulas and linking the bivariate copulas to a graph theoretical object, called \textit{regular vine} \parencite{Aas2009, Bedford2001}. This allows a more flexible dependence modeling. 
    \item Similar to \textcites{Vatter2015, Vatter2018}, we employ conditional bivariate copulas, where the copula parameter is derived from Kendall's $\tau$ to which the covariates  are linked in terms of a generalized linear model (GLM, \cite{Nelder1972}).
    \item We use the gradient-boosting approach based on \textcite{Buehlmann2007} for the conditional bivariate copula estimation. Moreover, we add an additional deselection step based on the attributable risk as a measure of variable importance \parencite{Stroemer2021}.
    \item To cover lots of possible dependency patterns, we provide besides of the bivariate Gaussian copula also rotation extended versions of the bivariate (survival) Clayton and (survival) Gumbel copula.
\end{compactenum}

In the following, we restrict us to the case of continuous responses, as for discrete or mixed continuous-discrete responses, e.g. adaptations of the copula families need to be made, which is beyond of the scope of this paper.
To the best of the authors knowledge, they are the first to combine conditional vine copulas with a gradient-boosting framework to handle low- and high-dimensional covariate settings. Furthermore, as there currently exits no software for a gradient-boosted based conditional vine copula, the authors provide an \texttt{R}-package called \texttt{boostCopula} on \texttt{\href{https://github.com/jobstdavid/boostCopula}{https://github.com/jobstdavid/boostCopula}}. 

The rest of the article is organized as follows: Section \ref{sec: Methodology} briefly introduces vine copulas and the conditional bivariate copulas in the gradient-boosting framework. Additionally, the estimation procedure for conditional vine copula is explained. A large simulation study for conditional vine copulas is conducted in Section \ref{sec: Simulations}, where we evaluate the effect estimates, the variable selection and copula family selection process in a low- and high-dimensional covariate setting. In Section \ref{sec: Application} we apply our new approach to the multivariate postprocessing of ensemble weather forecasts and present the results. We close with a conclusion and outlook in Section \ref{sec: Conclusion and outlook}.

\section{Methodology}
\label{sec: Methodology}

\subsection{Vine copulas}

Following Sklar's Theorem \parencite{Sklar1959}, the density function $f$ of an absolutely continuous $d$-dimensional distribution function $F\sim \bm{Y}=(Y_1,\ldots, Y_d)$ with marginal distributions $F_1, \ldots, F_d$ of $Y_1,\ldots, Y_d$ and corresponding copula $C$ is given by 
\begin{align}
	f(y_1,\ldots, y_d)=c(F_1(y_1), \ldots, F_d(y_d))\cdot \prod\limits_{i=1}^{d}f_i(y_i),
\end{align}
where $f, f_1, \ldots, f_d, c$ are the density functions of $F, F_1,\ldots, F_d, C$ respectively. To allow for a more flexible dependence modeling, \textcite{Bedford2001, Bedford2002} provided a decomposition of the copula density $c$ in $d(d-1)/2$ bivariate copula densities. As such a decomposition is not unique, it can be structured by a graphical model called \textit{regular vine} (R-vine) which consists of a sequence of linked trees $T_i=(N_i, E_i)$ with node set $N_i$ and edge set $E_i$ for $i=1,\ldots, d-1$, where the edges in one tree become the nodes of the next one. By identifying each edge $e\in E_i$ in the tree sequence with a bivariate copula density (\textit{pair-copula density}) $c_{a_e,b_e;D_e}$, the vine copula density $c$ can be written as the product of all pair-copula densities
\begin{align}
	c(\bm{u})=\prod\limits_{i=1}^{d-1}\prod\limits_{e\in E_i}^{}c_{a_e, b_e;D_e}(u_{a_{e}\vert D_{e}}, u_{b_{e}\vert D_{e}}; \bm{u}_{D_{e}}),
 \label{eq: r-vine copula density}
\end{align} 
where $\bm{u}:=(u_1,\ldots, u_d)\in [0,1]^d$ is the realization of $\bm{U}:=(U_1,\ldots,U_d)=(F_1(Y_1), \ldots, F_d(Y_d))$, $u_{k_e\vert D_e}:=C_{k_e\vert D_e}(u_{k_e}\vert \bm{u}_{D_e})$ is the conditional distribution of $U_{k_e}\vert \bm{U}_{D_e}$ which can be calculated recursively \parencite{Joe1996a} and $\bm{u}_{D_e}:=(u_l)_{l\in D_e}$ is a sub-vector of $\bm{u}$. A decomposition of a density as in Equation \eqref{eq: r-vine copula density} is therefore called \textit{pair-copula decomposition} of the joint density $c$. The set $D_e$ denotes the \textit{conditioning set} and the singletons $a_e, b_e$ are called  \textit{conditioned sets}. Consequently, the nodes labeled by $1,\ldots, d$ in the first tree $T_1$ correspond to the variables $U_1,\ldots,U_d$ and the edges $e$ labeled as $(a_e,b_e;D_e)=(a_e,b_e)$ stand for the unconditional dependence of variables $U_{a_e}$ and $U_{b_e}$ modeled by the copula density $c_{a_e,b_e}$. In the subsequent trees the edges labeled as $(a_e,b_e;D_e)$ describe the dependence of $U_{a_e}$ and $U_{b_e}$ conditioned on $\bm{U}_{D_e}=\bm{u}_{D_e}$ associated to the copula density $c_{a_e,b_e;D_e}$. 
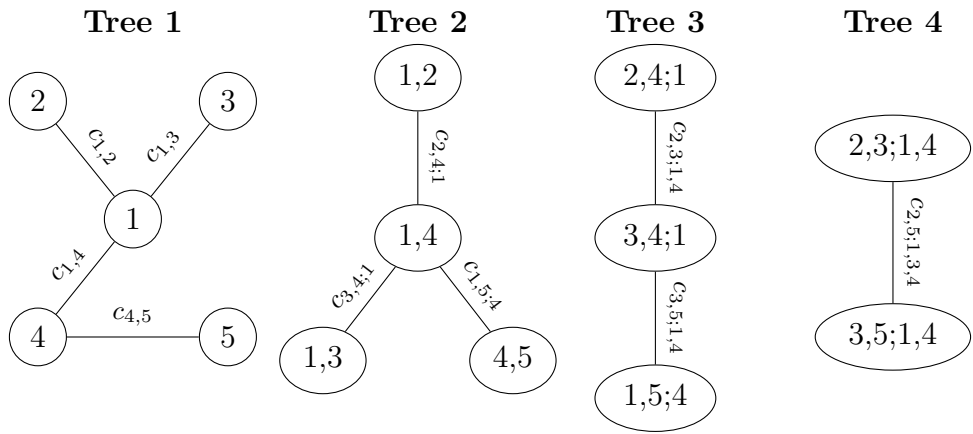
\begin{figure}[h!]
\begin{center}
\begin{tikzpicture}[scale = 1.25]
\usetikzlibrary{shapes}

	\node (Tree 1) at (-6, 0.6) {\textbf{Tree 1}};
	\node (Tree 2) at (-3, 0.6) {\textbf{Tree 2}};
	\node (Tree 3) at (-0.5, 0.6) {\textbf{Tree 3}};
	\node (Tree 4) at (2, 0.6) {\textbf{Tree 4}};

      \node[ellipse, minimum size=0.75cm, draw=black] (2) at (-7,-0.25) {2};
      \node[ellipse, minimum size=0.75cm, draw=black] (3) at (-5, -0.25) {3};
      \node[ellipse, minimum size=0.75cm, draw=black] (1) at (-6, -1.5) {1};
      \node[ellipse, minimum size=0.75cm, draw=black] (4) at (-7,-2.75) {4};
      \node[ellipse, minimum size=0.75cm, draw=black] (5) at (-5,-2.75) {5};

	\draw (2)--node[midway,above,sloped]{\footnotesize $c_{1,2}$}(1)
			(3)--node[midway,above,sloped]{\footnotesize $c_{1,3}$}(1)
			(1)--node[midway,above,sloped]{\footnotesize $c_{1,4}$}(4)
			(4)--node[midway,above,sloped]{\footnotesize $c_{4,5}$}(5);


      \node[ellipse, minimum size=0.75cm, draw=black] (12) at (-3,0) {1,2};
      \node[ellipse, minimum size=0.75cm, draw=black] (14) at (-3, -1.7) {1,4};
      \node[ellipse, minimum size=0.75cm, draw=black] (13) at (-4, -3) {1,3};
      \node[ellipse, minimum size=0.75cm, draw=black] (45) at (-2, -3) {4,5};		
      
      	\draw (12)--node[midway,above,sloped]{\footnotesize $c_{2,4;1}$}(14)
			(14)--node[midway,above,sloped]{\footnotesize $c_{3,4;1}$}(13)
			(14)--node[midway,above,sloped]{\footnotesize $c_{1,5;4}$}(45);


      \node[ellipse, minimum size=0.75cm, draw=black] (421) at (-0.5, 0) {2,4;1};
      \node[ellipse, minimum size=0.75cm, draw=black] (431) at (-0.5, -1.7) {3,4;1};
      \node[ellipse, minimum size=0.75cm, draw=black] (154) at (-0.5, -3.4) {1,5;4};	
      
      	\draw (421)--node[midway,above,sloped]{\footnotesize $c_{2,3;1,4}$}(431)
			(431)--node[midway,above,sloped]{\footnotesize $c_{3,5;1,4}$}(154);
			
			
      \node[ellipse, minimum size=0.75cm, draw=black] (3514) at (2, -0.75) {2,3;1,4};
      \node[ellipse, minimum size=0.75cm, draw=black] (3214) at (2, -2.75) {3,5;1,4};
      
      	\draw (3214)--node[midway,above,sloped]{\footnotesize $c_{2,5;1,3,4}$}(3514);	
      	
			      	
\end{tikzpicture}	
\end{center}
	\caption{5-dimensional regular vine and corresponding pair-copula densities.}
	\label{fig: R-vine}
\end{figure}
However, in the theory of vine copulas it is common to agree on the simplifying assumption \parencite{Stoeber2013} to allow for a tractable estimation of the pair-copula densities. Specifically, this means that we ignore the influence of the variables in the conditioning set $\bm{u}_{D_e}$ of the pair-copulas, i.e. $c_{a_e, b_e;D_e}(u_{a_{e}\vert D_{e}}, u_{b_{e}\vert D_{e}}; \bm{u}_{D_{e}})= c_{a_e, b_e;D_e}(u_{a_{e}\vert D_{e}}, u_{b_{e}\vert D_{e}})$ for all edges $e$ in a regular vine. Consequently, the pair-copula densities $c_{a_e, b_e;D_e}$ describe a partial dependence, rather than a conditional dependence. The conditioning on $\bm{u}_{D_e}$ is solely captured by its arguments $u_{a_{e}\vert D_{e}}$ and $u_{a_{e}\vert D_{e}}$. Therefore, the vine copula density in Equation \eqref{eq: r-vine copula density} simplifies to
\begin{align}
	c(\bm{u})=\prod\limits_{i=1}^{d-1}\prod\limits_{e\in E_i}^{}c_{a_e, b_e;D_e}(u_{a_{e}\vert D_{e}}, u_{b_{e}\vert D_{e}}).
 \label{eq: simplified r-vine copula density}
\end{align}

In Figure \ref{fig: R-vine} a 5-dimensional regular vine is visualized with corresponding (simplified) vine copula density given by
\begin{align}
	c(u_1, \ldots, u_5)&=c_{1,2}(u_1,u_2)\cdot c_{1,3}(u_1,u_3)\cdot c_{1,4}(u_1,u_4)\cdot c_{4,5}(u_4, u_5)\tag*{Tree 1}\\
	&\cdot c_{2,4;1}(u_{2\vert 1}, u_{4\vert 1})\cdot c_{3,4;1}(u_{3\vert 1}, u_{4\vert 1})\cdot c_{1,5;4}(u_{1\vert 4}, u_{5\vert 4})\tag*{Tree 2}\\
	&\cdot c_{2,3;1,4}(u_{2\vert 1,4}, u_{3\vert 1,4})\cdot c_{3,5;1,4}(u_{3\vert 1,4}, u_{5\vert 1,4})\tag*{Tree 3}\\
	&\cdot c_{2,5;1,3,4}(u_{2\vert 1,3,4}, u_{5\vert 1,3,4})\tag*{Tree 4}.
\end{align}

\subsection{Conditional vine copulas}

To obtain a conditional vine copula, the parameters of the bivariate copulas need to depend on covariates $\bm{Z}:=(Z_0,Z_1,\ldots, Z_p)$ with realizations $\bm{z}:=(z_0, z_1,\ldots, z_p)\in \R^{p+1}$, i.e. $c_{a_e, b_e;D_e}(u_{a_{e}\vert D_{e}}, u_{b_{e}\vert D_{e}}; \theta_{a_e, b_e;D_e}(\bm{z}))$
where $\theta_{a_e, b_e;D_e}(\bm{z})\in \Theta_{a_e, b_e;D_e}$ denotes the copula parameter depending on $\bm{z}$ for edge $e=(a_e,b_e;D_e)$ in the regular vine and $\Theta_{a_e, b_e;D_e}$ is the corresponding parameter space. Consequently, a conditional vine copula density is given by 
\begin{align}
	c(\bm{u};\bm{z})=\prod\limits_{i=1}^{d-1}\prod\limits_{e\in E_i}^{}c_{a_e, b_e;D_e}(u_{a_{e}\vert D_{e}}, u_{b_{e}\vert D_{e}}; \theta_{a_e, b_e;D_e}(\bm{z})).
 \label{eq: cond_vine_cop}
\end{align} 

\subsection{Gradient-boosted conditional bivariate copulas}
\label{sec: Gradient-boosted conditional bivariate copulas}

In the following, we describe the model setup for the gradient-boosted conditional bivariate copulas densities $c(u_1,u_2;\theta(\bm{z}))$,
which are the building blocks of the resulting conditional vine copula density, see Equation \eqref{eq: cond_vine_cop}. In Section \ref{sec: Estimation of conditional vine copulas} we then return to the estimation of the resulting conditional vine copula.

\subsubsection{Conditional bivariate copulas}

As a bijective transformation $h:(-1,1)\to \Theta$ between the Kendall's $\tau\in(-1,1)$ and a copula parameter $\theta\in \Theta$ with corresponding parameter space $\Theta\subseteq \R$ for common copula families can be derived (see Table \ref{tab: par_trafo}), we model the Kendall's $\tau$ correlation coefficient by covariates. Specifically, we assume a generalized linear model (GLM, \cite{Nelder1972}) with linear predictor
\begin{align}\eta(\bm{z}):=\bm{\beta}\bm{z}^T=\beta_0z_0+\beta_1z_1+\ldots+\beta_p z_p,
	\label{eq: glm}
\end{align}
and parameters $\bm{\beta}:=(\beta_0, \beta_1,\ldots, \beta_p)\in \R^{p+1}$. Note, that an intercept in the GLM can be obtained by forcing $Z_0=1$.

\begin{table}[h!]
\begin{center}
\begin{tabular}{c | c c c} 
\toprule
Family & Gaussian & Clayton I, II, & Gumbel I, II\\  \hline
$h:(-1,1)\to \Theta\subseteq\R$ & 
$h(t)=\sin\left(\frac{\pi}{2}t\right)$ & 
$h(t)=\frac{2t}{1-\abs{t}}$ &
$h(t)=\frac{\mathrm{sign}(t)}{1-\abs{t}}$ \\
\bottomrule
\end{tabular}
\end{center}
\caption{Bijective transformations $h$ for the considered copula families.}
\label{tab: par_trafo}
\end{table}

To ensure the restrictions on the Kendall's $\tau$ correlation coefficient of the considered copula families, we use the Fisher transformation $f:\R \to (-1,1),\quad f(\eta):=\tanh(\eta)$
to map the linear predictor $\eta$ to the range of the Kendall's $\tau$ correlation coefficient, where we assume the link to be correctly specified. Summing up, we model a single copula parameter via 
\begin{align}
	 \theta(\bm{z}):=h(f(\eta(\bm{z}))),
\end{align}
where we first transform the linear predictor $\eta$ via the link function $f$ to the Kendall's $\tau$ coefficient range and the result thereof into the copula parameter domain via the bijection $h$. This approach is in contrast to \textcite{Hans2022}, who modeled the copula parameter directly. However, the parameterization of Kendall's $\tau$ correlation coefficient has a simpler interpretation than a copula parameter, especially when the copula parameters live in different domains. 

\begin{figure}[h!]
	\begin{center}
		\includegraphics[scale = 0.4]{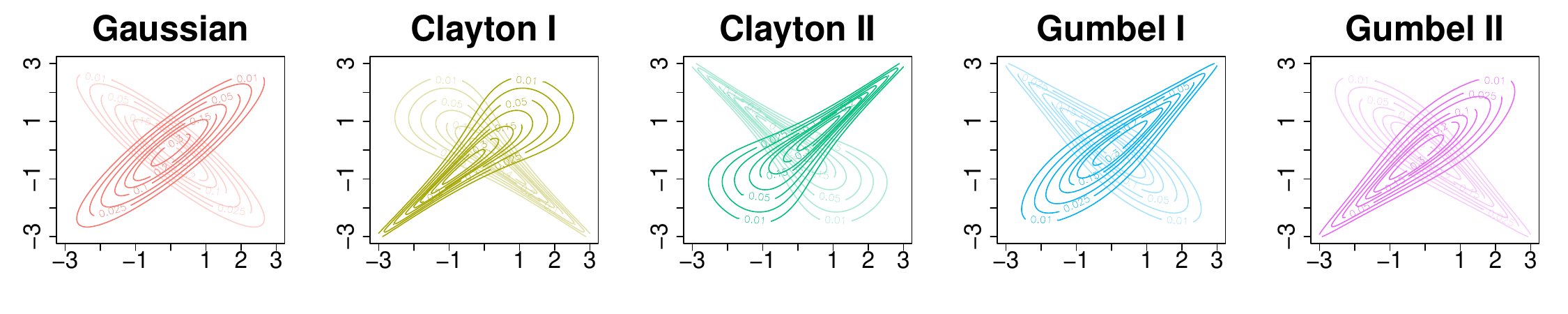}
	\end{center}	
	\caption{Density contour plots of the Gaussian, Clayton I, Clayton II, Gumbel I, Gumbel II copula for $\tau=0.7$ (dark) and $\tau = -0.7$ (bright).}
	\label{fig: bicop_contour}
\end{figure}

For each conditional bivariate copula in the conditional vine copula the same set of one-parametric copula families is allowed, from which the best fitting copula family is chosen (see Section \ref{sec: Copula family selection}). The set of all available copula families contains the Gaussian copula from the Elliptical copula class. However, the occurrence of tail dependence \parencite{McNeil2005} supports the investigation of copula families being different from Gaussian. Therefore, we additionally consider the Clayton and Gumbel copula from the Archimedean copula class which can capture lower and upper tail-dependence, respectively. 
We extend the Clayton and Gumbel copula by their 90 degrees counter-clockwise rotations to handle the negative Kendall's $\tau$ correlation coefficient as well, via 
\begin{align}
c_{\text{Family I}}(u_1,u_2;\tau):=\begin{cases}
c_{\text{Family}}(u_1,u_2;\tau),& \tau \geq 0,\\
c_{\text{Family}}(u_2,1-u_1;-\tau),& \tau<0,
\end{cases}	
\end{align}
for $\text{Family}\in \mng{\text{Clayton}, \text{Gumbel}}$. To obtain even richer dependence structures, survival copulas for Clayton I and Gumbel I are introduced which are 180 degrees counter-clockwise rotated versions thereof, i.e. 
\begin{align}
c_{\text{Family II}}(u_1,u_2;\tau):=c_{\text{Family I}}(1-u_1,1-u_2;\tau),
\end{align}
for $\text{Family}\in \mng{\text{Clayton}, \text{Gumbel}}$. This yields in total the 5 different copula families Gaussian, Gumbel I, Gumbel II, Clayton I, Clayton II visualized in Figure \ref{fig: bicop_contour} from which the best fitting one can be chosen. Note, that we are actually able to cover 9 different copula families by estimating only the previously mentioned 5 copula families which saves computational resources, too.

\subsubsection{Estimation via gradient-boosting}
\label{sec: Estimation via gradient-boosting}

The estimation of the linear predictor $\eta$ corresponding to the conditional bivariate copulas is based on a gradient-boosting approach. For a pair of standard uniformly distributed random variables $(U_1,U_2)$ with realization $(u_1,u_2)\in [0,1]^2$ the associated optimization problem can be written as
\begin{align}
	\widehat{\eta}_{\theta}=\argmin\limits_{\eta}\mathbb{E}_{U_1, U_2\vert \bm{Z}}[\ell(U_1,U_2; \eta(\bm{Z}))],
\end{align}

where $\ell$ denotes the loss function. In case of a data set with $N$ observations, the (conditional) expectation $\mathbb{E}_{U_1, U_2\vert \bm{Z}}$ is replaced by the empirical risk
\begin{align}
	\frac{1}{N}\sum\limits_{i=1}^{N}\ell\big(u_{1}^{(i)},u_{2}^{(i)};\eta(\bm{z}^{(i)})\big),
\end{align}

measuring how well the (transformed) linear predictor fits the Kendall's $\tau$ correlation coefficient inherent in the copula data. Furthermore, we select the negative log-likelihood of the conditional bivariate copula density as loss function $\ell$.
\algnewcommand\algorithmicinit{\textbf{Initialization}}
\algnewcommand\init{\item[\algorithmicinit]}

\algnewcommand\algorithmicboost{\textbf{Boosting}}
\algnewcommand\boost{\item[\algorithmicboost]}

\algnewcommand\algorithmicfinal{\textbf{Finalization}}
\algnewcommand\final{\item[\algorithmicfinal]}

\begin{algorithm}[h!]
\caption{Gradient-boosting}\label{alg: Gradient-boosting}
\begin{algorithmic}

\init  Initialize $(\beta_0^{[0]}, \beta_1^{[0]},\ldots,\beta_p^{[0]}):=(0,\ldots,0)$.

\boost For each iteration $m = 1,\ldots, m_{\text{stop}}$:
\State	1. Calculate the negative gradient $\bm{g}\in \R^N$ of the loss function:

$$\bm{g}:=(g^{(i)})_{i=1,\ldots,N}:=-\left(\frac{\partial}{\partial \eta}\ell\big(u_{1}^{(i)}, u_{2}^{(i)}; \eta^{[m-1]}(\bm{z}^{(i)})\big)\right)_{i=1,\ldots, N}.$$
\State 2. Fit for each covariate a separate linear regression model without intercept in the sense of ordinary least squares to the negative gradient:
$$g^{(i)}\left(z_j^{(i)}\right)=\widehat{\beta}_jz_j^{(i)},\quad i=1,\ldots,N,$$
with slope $\widehat{\beta}_j\in \R$ for each covariate $Z_j$, $j=0,1,\ldots,p$.

\State 3. Select the index $j^\ast$ which minimizes the residual sum of squares criterion:
$$j^\ast:=\argmin\limits_{j=0,1,\ldots,p} \; \sum\limits_{i=1}^{N}\left(g^{(i)}-\widehat{\beta}_jz_{j}^{(i)}\right)^2.$$
\State 4. Update the current estimate by
$$\beta_{j^\ast}^{[m]}=\beta_{j^\ast}^{[m-1]}+\nu\cdot \widehat{\beta}_{j^\ast},\qquad \beta_{j}^{[m]}=\beta_{j}^{[m-1]},\quad j\in \mng{0,1,\ldots,p}\backslash\mng{j^\ast},$$
\State using step length $\nu\in (0,1]$.
\final Set $(\beta_0,\beta_1,\ldots,\beta_p):=(\beta_0^{[m_{\text{stop}}]}, \beta_1^{[m_{\text{stop}}]}, \ldots, \beta_p^{[m_{\text{stop}}]})$.
\end{algorithmic}
\label{alg: boosting}
\end{algorithm} 
The basic idea of gradient-boosting is to minimize the empirical risk by a stepwise gradient-descent of the loss function. The procedure \parencite{Buehlmann2003, Buehlmann2007} is summarized in Algorithm \ref{alg: boosting}. Consequently, due to the selection of a single covariate and the update of the corresponding coefficient (step 3. and 4. of Algorithm \ref{alg: boosting}) in each iteration, the algorithm performs a variable selection. This procedure is carried out until a maximum number of iterations $m_{\text{stop}}$ is reached \parencite{Hofner2012}. The maximum number of boosting iterations $m_{\text{stop}}$ is an important tuning parameter for the gradient-boosting procedure as it helps to prevent overfitting and supports the sparsity of the final model by a data-driven variable selection and yields to an improved prediction accuracy \parencite{Hofner2012}. It is usually selected by cross-validation (CV) or AIC minimization. Another tuning parameter is the step length $\nu$ for which a small value of e.g. $\nu=0.1$ has been established as appropriate in most cases \parencite{Hofner2012}.

\subsubsection{Deselection of covariates}

Especially for low-dimensional settings, the gradient-boosting algorithm in Section \ref{sec: Estimation via gradient-boosting} tends to include too many covariates, which results in a slow overfitting. To overcome this issue and to achieve sparse models, we follow \textcite{Stroemer2021} and apply a deselection procedure for the covariates after the gradient-boosted estimation of the conditional bivariate copula. In a subsequent step, we boost the conditional bivariate copula again only with the remaining covariates from the deselection procedure using the already previously determined optimal number of boosting iterations $m_{\text{opt}}$ (by cross-validation or AIC minimization). For the deselection step, the attributable risk reduction 

\begin{align}
R_j:=\sum\limits_{m=1}^{m_{\text{stop}}}\mathds{1}\mng{j=j^{\ast[m]}}\cdot (r^{[m-1]}-r^{[m]}),\quad j=0,1,\ldots, p,
\end{align}

is considered as a measure for variable importance of the $j$-th covariate $Z_{j}$.  

\begin{algorithm}[h!]
\caption{Gradient-boosting with additional deselection of covariates}\label{alg: complete gradient-boosting}
\begin{algorithmic}

\State \textbf{1. Initial boosting:} Estimate the conditional bivariate copula via gradient-boosting with an initial number of boosting iterations $m_{\text{stop}}$.
\State \textbf{2. Early stopping:} Tune the optimal number of boosting iterations $m_{\text{opt}}$ via cross-validation or AIC minimization.
\State \textbf{3. Deselection of covariates:} Deselect the covariates with the smallest impact on the risk reduction according to Equation \eqref{eq: rr}.
\State \textbf{4. Final boosting:} Boost the conditional bivariate copula again with the remaining covariates of step 3. and $m_{\text{opt}}$ of step 2.  
\end{algorithmic}
\end{algorithm}

Here, $\mathds{1}$ denotes the indicator function, $j^{\ast[m]}$ corresponds to the selected covariate $Z_{j^{\ast[m]}}$ of the initial boosting and $r^{[m-1]}-r^{[m]}$ gives the risk reduction at iteration $m$ for the risk $r^{[m]}$ at iteration $m$ and the risk $r^{[m-1]}$ at iteration $m-1$. The risk is given by the value of loss function $\ell$, i.e. in our case the negative log-likelihood. Consequently, the $j$-th covariate $Z_j$ is deselected if 
\begin{align}
	R_j<\gamma \cdot (r^{[0]}-r^{[m_{\text{stop}}]}), \label{eq: rr}
\end{align}
for a given threshold $\gamma\in (0,1)$, where $r^{[0]}-r^{[m_{\text{stop}}]}$ is the total risk reduction. Consequently, only covariates for which the relative risk contribution is greater than or equal to the threshold $\gamma$ will be kept in the model. \textcite{Stroemer2021} suggest the use of a small threshold, e.g. $\gamma = 0.01$, but also outline, that the choice of $\gamma$ depends on the research situation. Therefore, the selection of $\gamma$ is a trade-off between more complex models with higher prediction accuracy and sparser, more interpretable models with possibly lower prediction performance \parencite{Stroemer2021}. 
Summing up, for a specified copula family the conditional bivariate copula is estimated in the four steps described in Algorithm \ref{alg: complete gradient-boosting}.

\subsubsection{Copula family selection} 
\label{sec: Copula family selection}


After having estimated a specified subset of copulas from the set of 5 copula families via Algorithm \ref{alg: complete gradient-boosting}, the copula family can be selected by goodness-of-fit measures, for example log-likelihood or complexity criteria, such as Akaike information criterion (AIC, \cite{Akaike1998}). As the gradient-boosting results in regularized model fits, the model complexity can be hard to evaluate \parencite{Hastie2007}. Therefore, \textcite{Hofner2016} suggest to use the predictive empirical risk, i.e. the evaluation of the negative log-likelihood on a new data set (out-of-sample). \textcite{Hans2022} employ this measure to decide which copula family should be chosen. Unfortunately, this comes with the drawback of a reduction in the size of the training data, which can lead to a loss in the prediction accuracy for the copula parameter. Furthermore, it might yield to a biased copula family selection, especially, when the training data size is not large. 
However, in the copula theory, the AIC is frequently applied for the copula family selection, as it has proven to be a reliable selection criterion \parencite{Brechmann2010}. Therefore, we investigate the AIC as copula selection criterion, where we approximately determine the degrees of freedom based on the number of nonzero coefficients (active set) in Equation \eqref{eq: glm}. According to \textcite{Buehlmann2007b}, the active set can work reasonable well as measure for the model degrees of freedom. 

\subsection{Estimation of conditional vine copulas}
\label{sec: Estimation of conditional vine copulas}

Until now, we have only introduced the estimation and selection of conditional bivariate copulas. In the following, we briefly describe the corresponding estimation procedure for a $d$-dimensional conditional vine copula with predetermined regular vine.

We use a sequential top-down estimation \parencites{Aas2009, Czado2019} for the conditional bivariate copulas in a conditional vine copula, where the conditional bivariate copulas are separately estimated via Algorithm \ref{alg: complete gradient-boosting} for the allowed set of copula families for each edge in tree $T_1$. Then, the copula family is selected as described in Section \ref{sec: Copula family selection} for each edge in tree $T_1$. Afterwards this estimation and selection procedure of the conditional bivariate copulas is applied to the remaining trees $T_j,\, j=2,\ldots, d-1$. While the copula data $\bm{u}=(u_1,\ldots,u_d)$ for the first tree $T_1$ is given, the copula data for the higher trees $T_j$ with $j=2,\ldots, d-1$ is unobserved. However, so-called \textit{pseudo-observations} calculated based on the recursion formula in \textcite{Joe1996a} can be used as copula data in the higher trees. Note that a regular vine might not always be given in advance and therefore needs to be estimated. However, as we do not consider this setting throughout the article, we refer to \textcites{Dissmann2013, Czado2019} for more details. Nonetheless, the regular vine estimation based on \textcite{Dissmann2013}  does not represent any obstacle for our suggested estimation procedure and is therefore implemented in the \texttt{R}-package \texttt{boostCopula} as well.

\section{Simulations}
\label{sec: Simulations}

In this section, we would like to analyze the properties of the gradient-boosted conditional vine copulas for low- and high-dimensional covariate settings. Similar to \textcite{Sanchez2024}, we carry out a simulation study where we first generate samples $(Z_1,\ldots,Z_{p-1})$ from a multivariate normal distribution $\mathcal{N}_{p-1}(0,\bm{\Sigma})$ with Toeplitz covariance matrix $\Sigma_{ij}=\rho^{\abs{i-j}}$ for $1\leq i,j\leq p-1$, correlation $\rho\in (0,1)$ and sample size $N$. Specifically, we consider two low-dimensional ($N=1000, p=101$; $N=2000, p=501$) and two high-dimensional ($N=1000, p=2001$; $N=2000, p=4001$) scenarios each in a low-correlated $\rho=0.2$ and high-correlated $\rho=0.8$ setting. For each case we perform $n=100$ simulation runs, where we assume for a standardized validation across all conditional bivariate copulas the linear predictor
\begin{align}
	\eta(\bm{z}):=0.1z_0-0.2z_1+0.3z_2+0.2z_3+0.5z_4-0.4z_5,
	\label{eq: sim_glm}
\end{align}
with intercept $Z_0=1$, meaning that the covariates $Z_0,Z_1,\ldots,Z_5$ are part of the true data generating process and thus informative, while the covariates $Z_6,\ldots,Z_{p-1}$ are not used to generate the data, and  thus can be considered as non-informative covariates. Afterwards, we generate samples from the conditional bivariate copula for a fixed copula family. For the estimation of the conditional bivariate copula we apply Algorithm \ref{alg: complete gradient-boosting} with maximum number of boosting iterations $m_{\text{stop}}=500$ and the commonly suggested specifications for the step length $\nu=0.1$ and the deselection threshold $\gamma = 0.01$.

In Section \ref{sec: Sim conditional bivariate copulas} we investigate the estimation and selection of all available 5 copula families (Gaussian, Gumbel I, Gumbel II, Clayton I, Clayton II) separately. In Section \ref{sec: Sim conditional vine copula} we examine the gradient-boosted estimation and selection of the conditional bivariate copulas for conditional vine copulas as described in Section \ref{sec: Estimation of conditional vine copulas}. Eventually, we would like to answer the following questions for Algorithm \ref{alg: complete gradient-boosting} by this simulation study:
\begin{enumerate}
	\item How well does Algorithm \ref{alg: complete gradient-boosting} estimate the coefficients for the informative covariates?
	\item Is Algorithm \ref{alg: complete gradient-boosting} able to select the informative and deselect the non-informative covariates?
	\item Are the specified copula families reasonably well selected according to AIC?
	\item Does AIC work sufficiently well as early stopping criterion in comparison to CV? 
	\item How do the results of questions 1.-3. look like for a  conditional vine copula depending on the tree level? 
\end{enumerate}

For the simulation study and the application in Section \ref{sec: Application} we use the software \texttt{R} \parencite{RCT2020}, running version 3.6.3 and the \texttt{R}-package \texttt{boostCopula}. Additionally, we run our codes on a cluster where we access 10 cores for code parallelization on each node.

\subsection{Conditional bivariate copulas}
\label{sec: Sim conditional bivariate copulas}

In this section, we investigate the estimation and selection of the five copula families Gaussian, Gumbel I, Gumbel II, Clayton I, Clayton II for low- and high-correlated covariates in low- and high-dimensional  settings. We compare the AIC and CV as stopping criteria as well analyze the performance of the AIC for the copula family selection.

\begin{figure}[h!]
	\begin{center}
		\includegraphics[scale = 0.4]{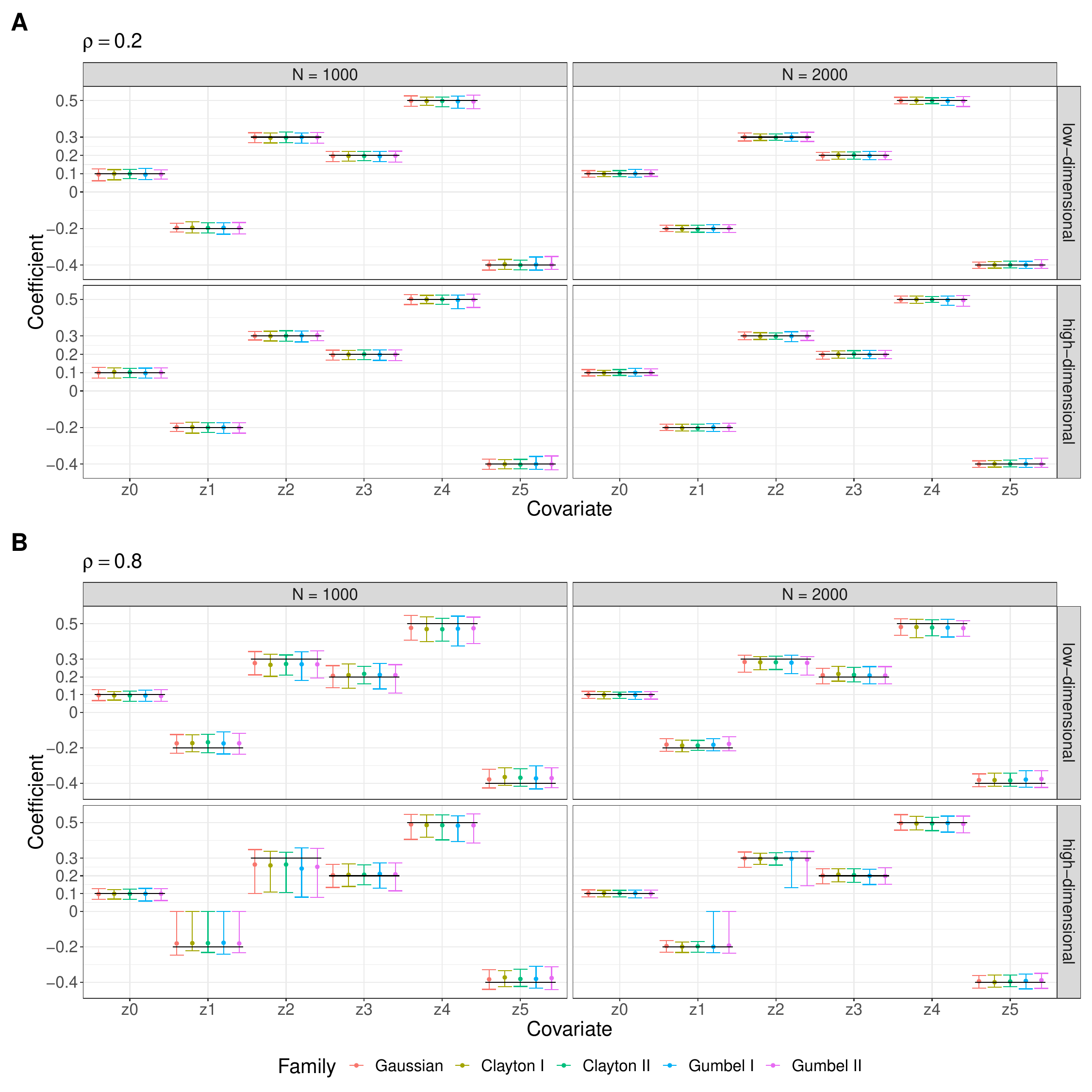}
	\end{center}	
	\caption{Summary of estimated coefficients with AIC as stopping criterion, where the bars represent the range from the 5\% to the 95\% quantiles and the points denote the median of the estimated coefficients for the low-correlated (A) and high-correlated (B) setting. The color represent the copula families. The horizontal black lines correspond to the true coefficients.}
	\label{fig: coef_bicop_loglik}
\end{figure}

\paragraph{Parameter estimation.} 
Figure \ref{fig: coef_bicop_loglik} presents the accuracy of the estimated coefficients for the informative covariates represented by the black lines of all considered copula families with AIC as stopping criterion. Firstly, we cannot observe large differences in the median coefficient estimates across the different copula families in all settings. Secondly, Algorithm \ref{alg: complete gradient-boosting} estimates the coefficients with less bias and variance in the low-correlated ($\rho=0.2$) setting, while  both of these properties become more pronounced for the high-correlated ($\rho=0.8$) setting. 
Furthermore, we can observe, that increasing the sample size from $N=1000$ to $N=2000$ yields to sharper and less biased coefficients independently of the correlation setting or covariate dimension. Eventually, Table \ref{tab: coef_bicop_medians} shows, that the AIC stopping criterion yields coefficient estimates which are in median closer to the true coefficients in contrast to the CV stopping criterion.

\begin{table}[h!]
\begin{center}
\resizebox{15cm}{!}{
\begin{tabular}{|c|c|c|c|c|c|c|c|} 
\hline
Dimension & Stopping criterion & $\beta_0=0.1$ & $\beta_1=-0.2$ & $\beta_2=0.3$ & $\beta_3=0.2$ & $\beta_4=0.5$ & $\beta_5=-0.4$ \\
\hline
\multirow{2}{2cm}{low-dimensional} & AIC & 0.098 & $-0.191$ & 0.292 & 0.202 & 0.492 & $-0.392$ \\ 
& CV & 0.093 & $-0.180$ & 0.282 & 0.200 & 0.482 & $-0.380$ \\ \hline 
\multirow{2}{2cm}{high-dimensional} & AIC & 0.100 & $-0.197$ & 0.297 & 0.201 & 0.497 & $-0.396$ \\ 
& CV & 0.088 & $-0.166$ & 0.270 & 0.198 & 0.471 & $-0.366$ \\ 
\hline
\end{tabular}
}
\caption{Medians of the estimated coefficients $\widehat{\beta}_0, \ldots, \widehat{\beta}_5$ for the informative covariates over all copula families, correlation settings $\rho \in \mng{0.2, 0.8}$ and sample sizes $N\in \mng{1000, 2000}$.}
\label{tab: coef_bicop_medians}
\end{center}
\end{table}

\begin{figure}[h!]
	\begin{center}
		\includegraphics[scale = 0.4]{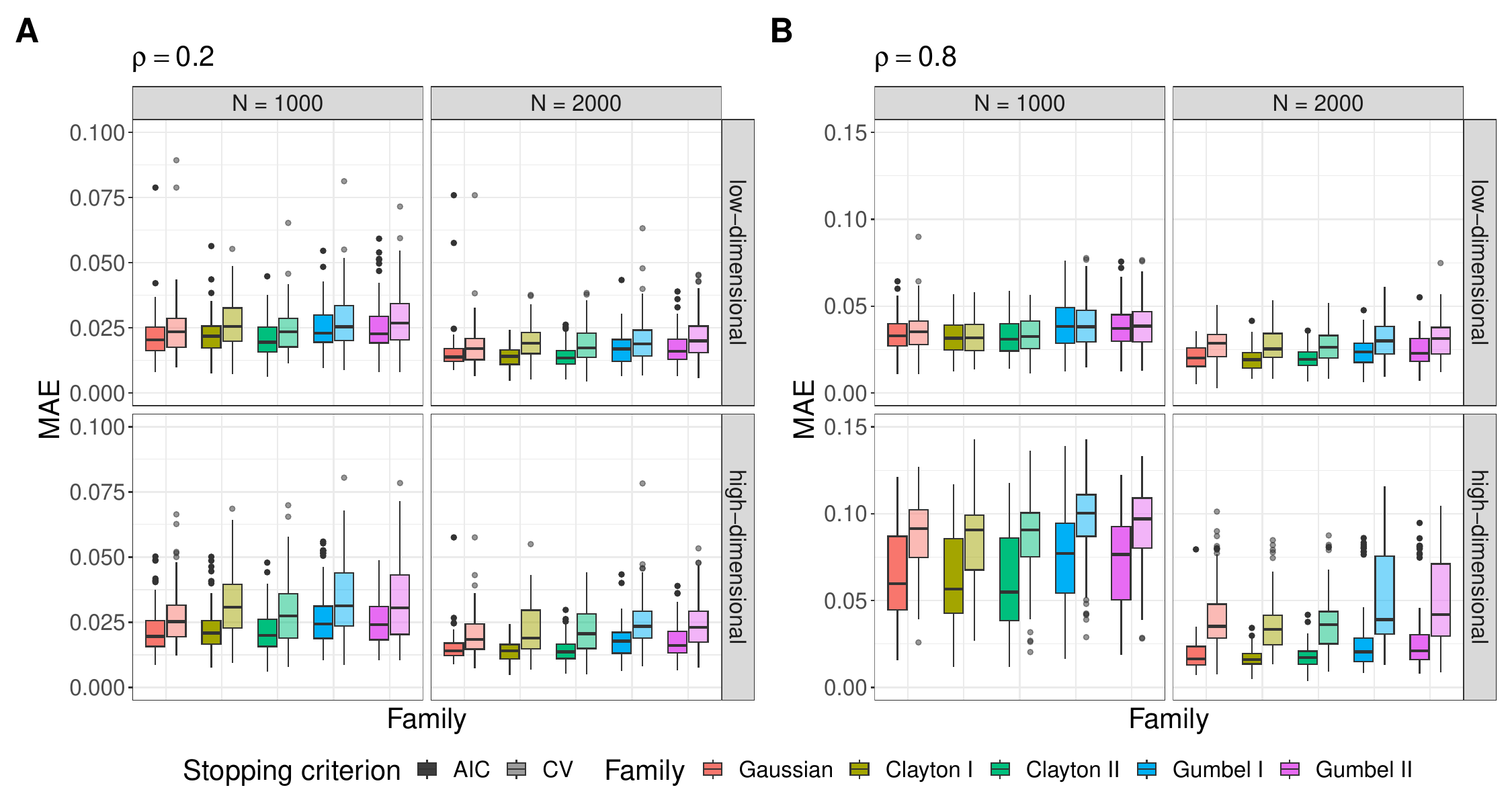}
	\end{center}	
	\caption{Boxplots of the mean absolute error (MAE) of Kendall's $\tau$ correlation coefficient for the low-correlated (A) and high-correlated (B) setting. The color represent the copula families while the color transparency indicates the stopping criterion.}
	\label{fig: mae_tau_bicop_aic}
\end{figure}

Latter observation is further illustrated in Figure \ref{fig: mae_tau_bicop_aic}, presenting the boxplots of the \textit{mean absolute errors} 
\begin{align}
	\mathrm{MAE}:=\frac{1}{N}\sum_{i=1}^{N}\vert h(f(\eta(\bm{z}^{(i)})))-h(f(\widehat{\eta}(\bm{z}^{(i)})))\vert,
\end{align}
for Kendall's $\tau$ correlation coefficient. Note, that one outlier (0.177) in the high-dimensional setting  $N=1000$, $\rho = 0.2$ for the Clayton II family using stopping criterion CV is not displayed for a better boxplot representation. The AIC stopping criterion yields in median to lower values for MAE than CV and increasing the sample size also helps to reduce the MAE in general. One reason for the superiority of AIC over CV is that AIC tends to stop a bit later than CV for the optimal number of boosting iterations $m_{\text{opt}}$ (see Appendix \ref{app: Conditional bivariate copulas}), which allows more optimization iterations for the coefficients after the deselection procedure than the earlier stopping CV. Furthermore, we can mostly detect no large differences among the copula families but observe slightly  higher MAE values in the high-correlated than in the low-correlated setting. Overall, these results are in line with the ones of Figure \ref{fig: coef_bicop_loglik} and Table \ref{tab: coef_bicop_medians}.

\paragraph{Covariate selection.}
\begin{figure}[h!]
	\begin{center}
		\includegraphics[scale = 0.4]{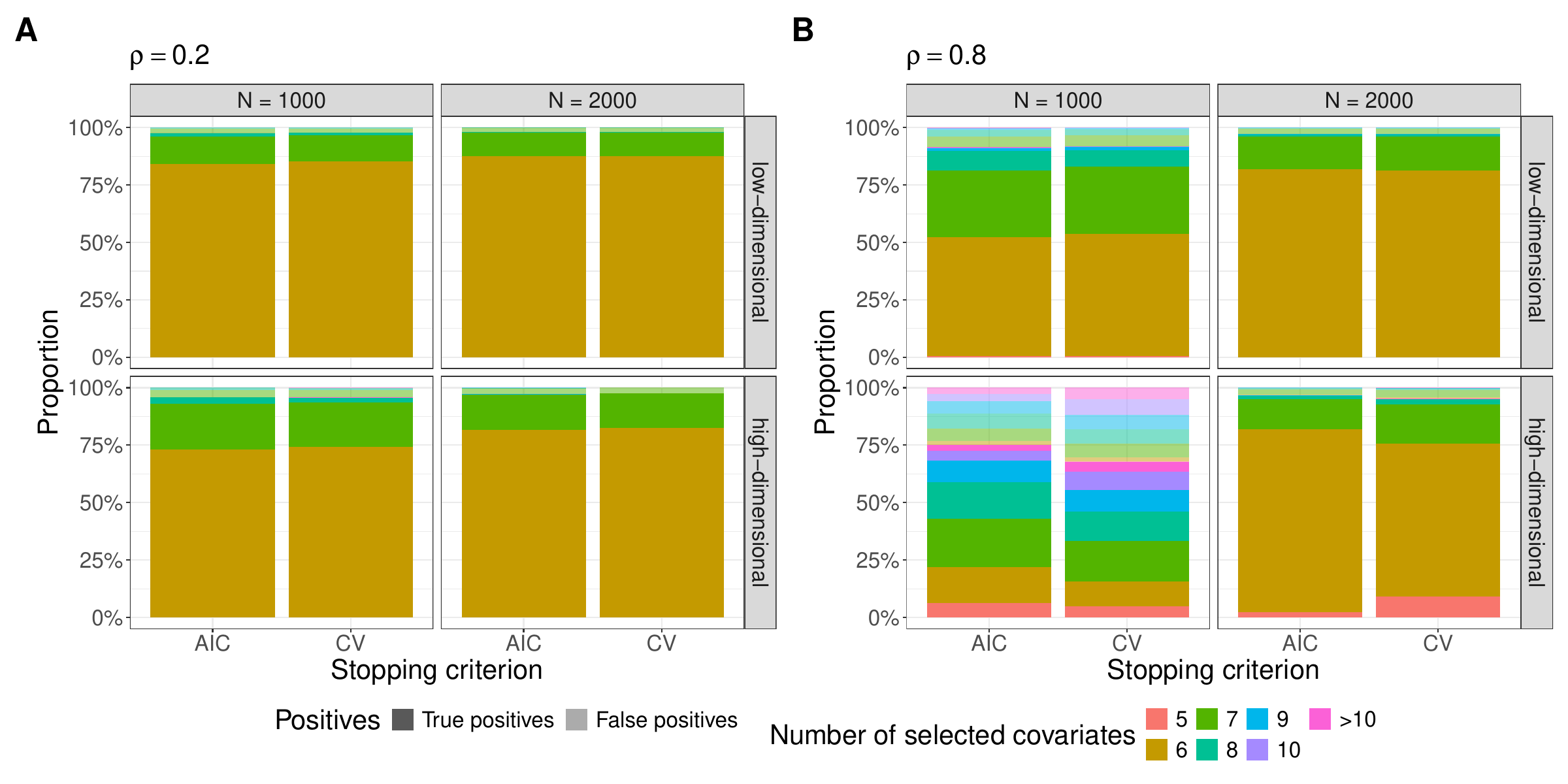}
	\end{center}	
	\caption{Bar charts for proportion of selected covariates across all 5 copula families and $n=100$ repetitions. The color transparency represents the proportion of true (dark) and false (light) positives for each number of selected covariates in the stacked bar charts.}
	\label{fig: tp_fp_bicop}
\end{figure}
As one might expect already from the previous results, there have been no strong differences among the copula families with respect to the covariate selection, and therefore we have a look at the covariate selection across all copula families in Figure \ref{fig: tp_fp_bicop}. The color transparency represents the proportion of true (dark) and false (light) positives for each number of selected covariates, while the sum of both proportions yield the total proportion for the selected number of covariates. In nearly each setting, six covariates are most frequently selected, followed by seven selected covariates. Furthermore, we observe that Algorithm \ref{alg: complete gradient-boosting} selects in over 70\% of all cases only the six truly informative covariates, except in the high-correlated settings with sample size $N=1000$. Especially in the high-correlated setting, we observe that increasing the sample size $N=2000$ yields to a strong improvement in the sense of favouring the  selection of all six informative covariates, reducing the spread in the number of selected covariates at the same time. With regard to the stopping criterion and proportion of true and false positives we can say, that there are no strong differences. With respect to the false positives, we deduce that its proportion compared to the true positives is rather small in all scenarios besides of the high-correlated, high-dimensional setting with sample size $N = 1000$.

\begin{figure}[h!]
	\begin{center}
		\includegraphics[scale = 0.4]{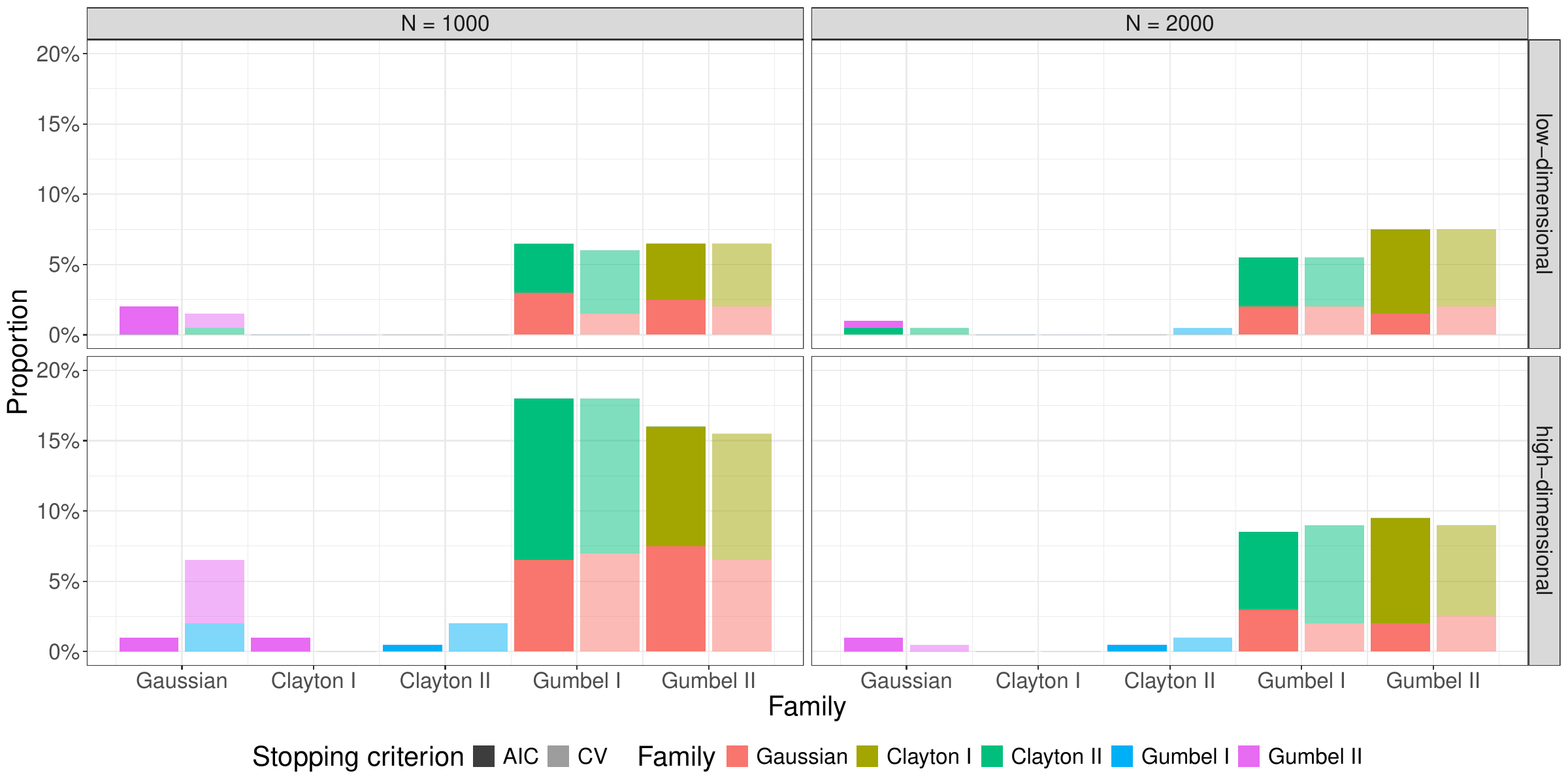}
	\end{center}	
	\caption{Percentage of false selected copula families for each specified copula family based on the AIC aggregated over the low-correlated and high-correlated setting. The colors represent the copula families while the color transparency indicates the stopping criterion.}
	\label{fig: fams_bicop_aic}
\end{figure}

\paragraph{Copula family selection.}
As there are no strong differences among the low- and high-correlated setting, Figure \ref{fig: fams_bicop_aic} shows the percentage of false selected copula families based on the AIC over all $n=100$ repetitions for both correlation settings. Firstly, the Gaussian, Clayton I, II families are identified overall the best, while instead of the Gumbel I, II families, other families are more often chosen. Depending on the given configuration, either AIC or CV as stopping criterion yields slightly higher selection rates for the underlying copula family, while increasing the sample size shows only strong effect on the selection rate for the high-dimensional configuration. Moreover, we conclude that the AIC works reasonably well in any of these settings to identify the true copula family. In comparison to other copula family selection criteria, e.g. (in-sample) log-likelihood or predictive risk we found that AIC performs comparable or sometimes even better (see Appendix \ref{app: Conditional bivariate copulas}).  

To sum up, we conclude that AIC is able to outperform CV as stopping criterion with respect to model accuracy, and it additionally provides a strong improvement with respect to the average computation times (see Appendix \ref{app: Conditional bivariate copulas}). Furthermore, the AIC is suitable to identify the true copula family. Therefore, we suggest to estimate the considered conditional bivariate copula families with AIC as stopping criterion and use the AIC as copula family selection criterion.

\subsection{Conditional vine copula}
\label{sec: Sim conditional vine copula}

Similar to \textcite{Vatter2018}, we choose a 5-dimensional vine copula as specified in Figure \ref{fig: R-vine}. The copula family for each pair-copula is randomly drawn from the available 5 copula families with equal probability. In a first scenario, called ``Selected'' model, the conditional bivariate copulas are estimated and selected as described in Section \ref{sec: Estimation of conditional vine copulas} with the sequential top-down estimation procedure. We apply Algorithm \ref{alg: complete gradient-boosting} with AIC as stopping criterion and the AIC as copula family selection criterion, as it was found to work well in Section \ref{sec: Sim conditional bivariate copulas}. In a second scenario, called ``Specified'' model, we only estimate the conditional bivariate copulas via gradient-boosting, while the copula families have already been selected. Additionally, only the informative covariates are included for the GLM and estimated via Algorithm \ref{alg: Gradient-boosting} without any covariate deselection afterwards. Latter setting serves as benchmark model to outline, how Algorithm \ref{alg: Gradient-boosting} performs under a correct model specification in contrast to the ``Selected'' model setting. For both scenarios we choose the same tuning parameters as in Section \ref{sec: Sim conditional bivariate copulas}.

\begin{figure}[h!]
	\begin{center}
		\includegraphics[scale = 0.4]{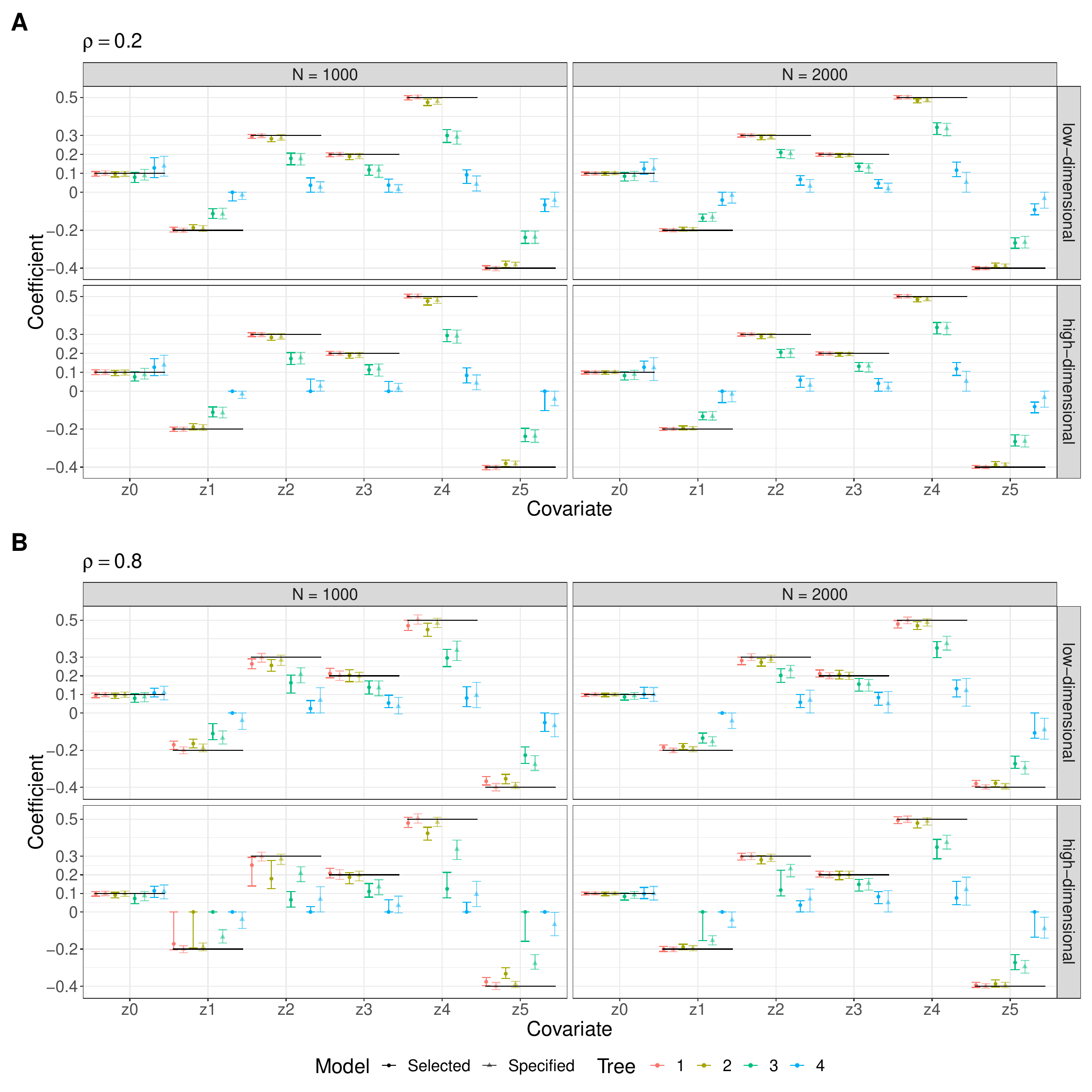}
	\end{center}	
	\caption{Summary of estimated coefficients, where the error bars represent the interquartile range and the points denote the median of the estimated coefficients for the low-correlated (A) and high-correlated (B) setting. The horizontal black lines correspond to the true coefficients. The colors represent the tree level while the color transparency indicates the copula model.}
	\label{fig: coef_rvine}
\end{figure}

\paragraph{Parameter estimation.}

Figure \ref{fig: coef_rvine} presents the accuracy of the estimated coefficients for the six informative covariates in each tree level. Independent of the considered setting, we observe that the true effects are estimated quite well in median for the first two tree levels, while the coefficients become biased towards a value of zero for tree level 3 and 4. This shrinkage towards zero has already been discovered for conditional vine copulas by \textcite{Vatter2018}, who determined the (penalized) maximum-likelihood coefficient estimates of the conditional bivariate copula employing a generalized ridge regression approach. Further investigations outlined that the Algorithm of \textcite{Vatter2018} showed a similar shrinkage towards zero for the coefficients in our simulation settings, even though this effect was not as pronounced as for the gradient-boosted based estimation (see Appendix \ref{app: Conditional vine copulas}). The bias towards zero in the coefficients for increasing tree levels is a result of the sequential estimation procedure, as the estimation errors in the current tree reproduce and hide the covariate effects in the subsequent trees. Although this shrinkage was not intended, it might motivate to specify even more parsimonious models with less parameters, e.g. by truncating the regular vine at a certain tree level.

Furthermore, we observe that the median coefficient estimates of the ``Specified'' models are a bit closer to the true coefficients in contrast to the ``Selected'' models. Additionally, increasing the sample size helps again to reduce the bias in the coefficient estimates. 

\begin{figure}[h!]
	\begin{center}
		\includegraphics[scale = 0.4]{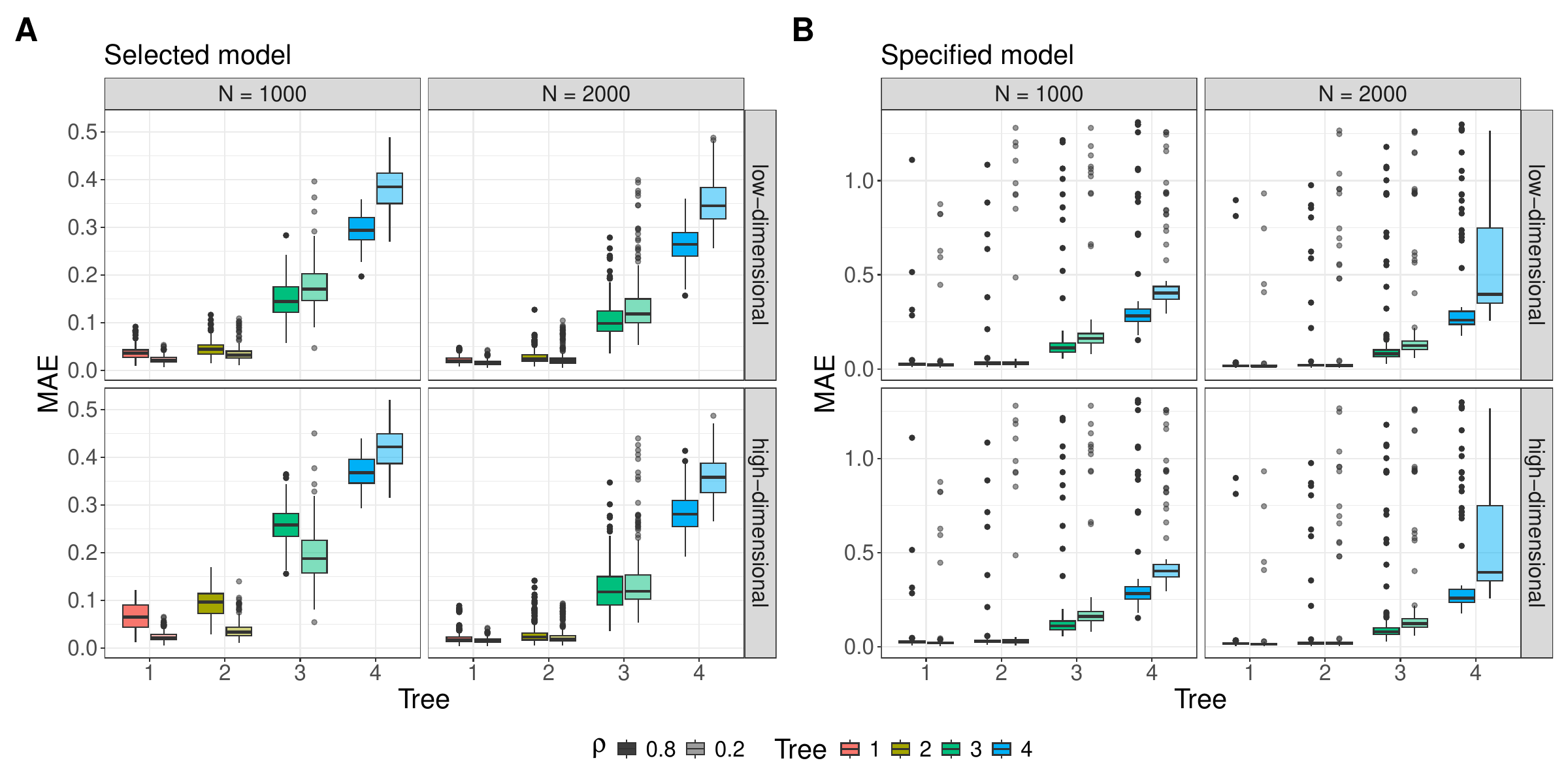}
	\end{center}	
	\caption{Boxplots of the mean absolute error (MAE) of Kendall's $\tau$ correlation coefficient for the selected (A) and specified (B) model. The colors represent the tree level while the color transparency stands for correlation $\rho$.}
	\label{fig: mae_tau_rvine}
\end{figure}

Figure \ref{fig: mae_tau_rvine} shows boxplots of the mean absolute error (MAE) of Kendall's $\tau$ correlation coefficient. We observe that the MAE becomes in general larger with increasing tree level, for each of the considered settings, which underlines our previous findings. While the median MAE is generally lower for the low-correlated setting in tree levels 1 and 2, the models in the high-correlated setting perform better with respect to the same measure in tree levels 3 and 4. Furthermore, there are no strong differences among the selected and specified model in terms of the median MAE, but we observe a large amount of outliers with high values for the specified model in each tree and simulation setting. These outliers can be traced back to the cases, where Algorithm \ref{alg: Gradient-boosting} selects in total only very few covariates, meaning specifically, a small number of the informative covariates (mostly one or two, see Appendix \ref{app: Conditional vine copulas}) for the ``Specified'' models. In contrast, Algorithm \ref{alg: complete gradient-boosting} chooses additional non-informative covariates in these situations, which leads to an overfitting for the ``Selected'' models, but lower MAE. Eventually, increasing the sample size reduces the MAE independently of the simulation setting. 

\newpage 

\paragraph{Covariate selection.}

\begin{figure}[h!]
	\begin{center}
		\includegraphics[scale = 0.4]{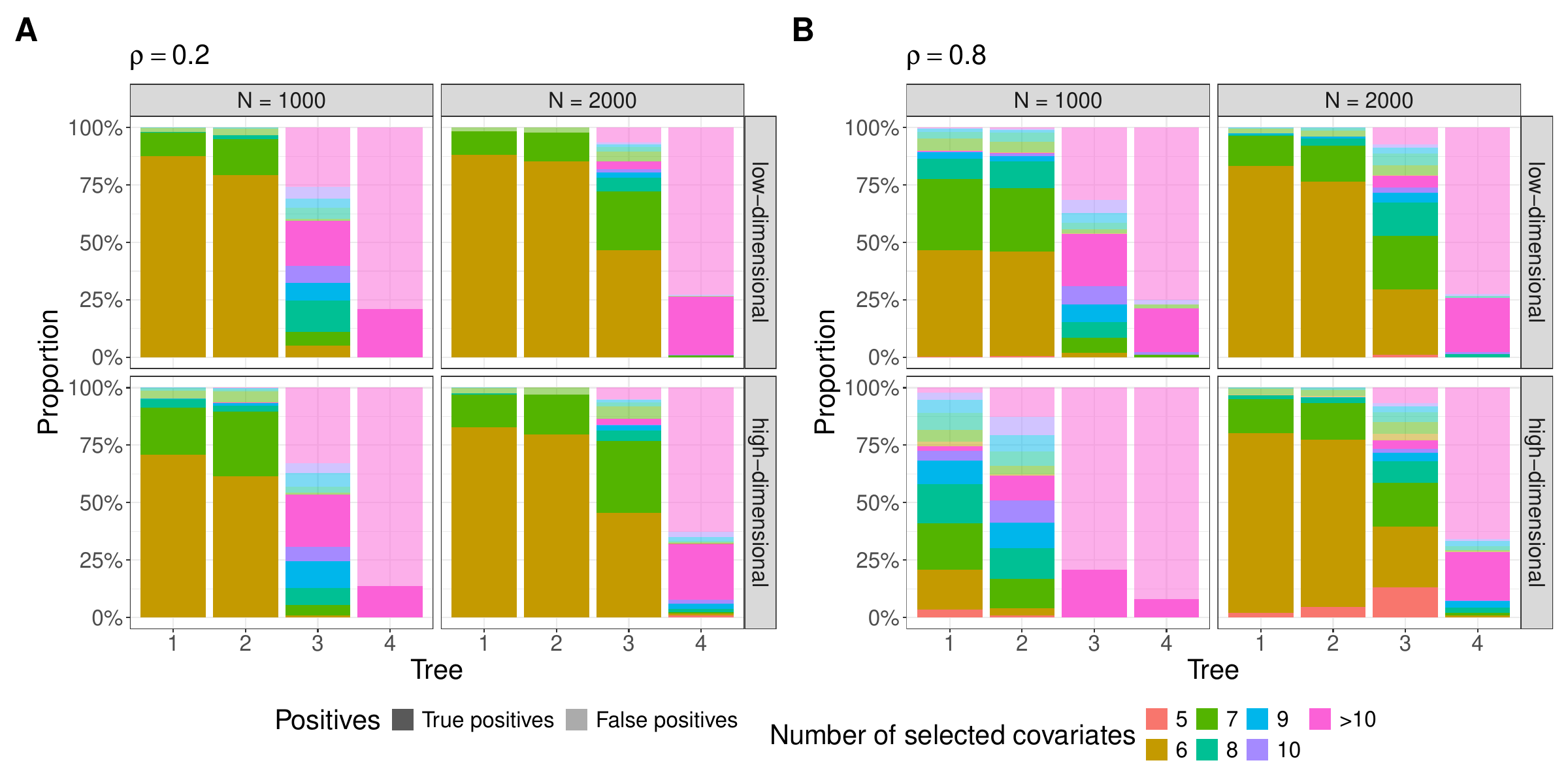}
	\end{center}	
	\caption{Bar charts for the proportion (\%) of selected covariates across all copulas for each tree level and $n=100$ repetitions in the ``Selected'' model for the low-correlated (A) and high-correlated (B) setting. The color transparency represents the proportion of true (dark) and false (light) positives for each number of selected covariates in the stacked bar charts.}
	\label{fig: tp_fp_rvine}
\end{figure}

In Figure \ref{fig: tp_fp_rvine}, the color transparency represents the proportion of true (dark) and false (light) positives for each number of selected covariates. Both values together yield the total proportion for each occurring number of selected covariates across all copulas in each tree level. For tree levels $1,2$, we observe that 6 or 7 covariates are most frequently selected, where an increased sample size helps to enhance the proportion of six selected informative covariates. Furthermore, the proportion of true positives is much higher than the proportion of false positives for the cases where 6 or 7 covariates are selected. Therefore we deduce that with increasing sample size it gets more likely, that all informative covariates are included in tree levels $1,2$ for the conditional bivariate copulas in the cases where 6 or 7 covariates are selected. With increasing tree level, there is a reduction in the proportion of cases where a moderate number (especially 6 or 7) of covariates is selected, while the proportion of cases where $8,9,10$ or more than $10$ covariates are selected clearly increases. For tree level 4, we observe that usually more than 10 covariates are selected. Furthermore, the proportion of false positives in comparison to true positives is clearly higher in tree level 4. In terms of the correlation setting, we only detect large differences in the proportion of cases for the numbers of selected covariates for the smaller sample size. 

\paragraph{Copula family selection.}
\begin{figure}[h!]
	\begin{center}
		\includegraphics[scale = 0.4]{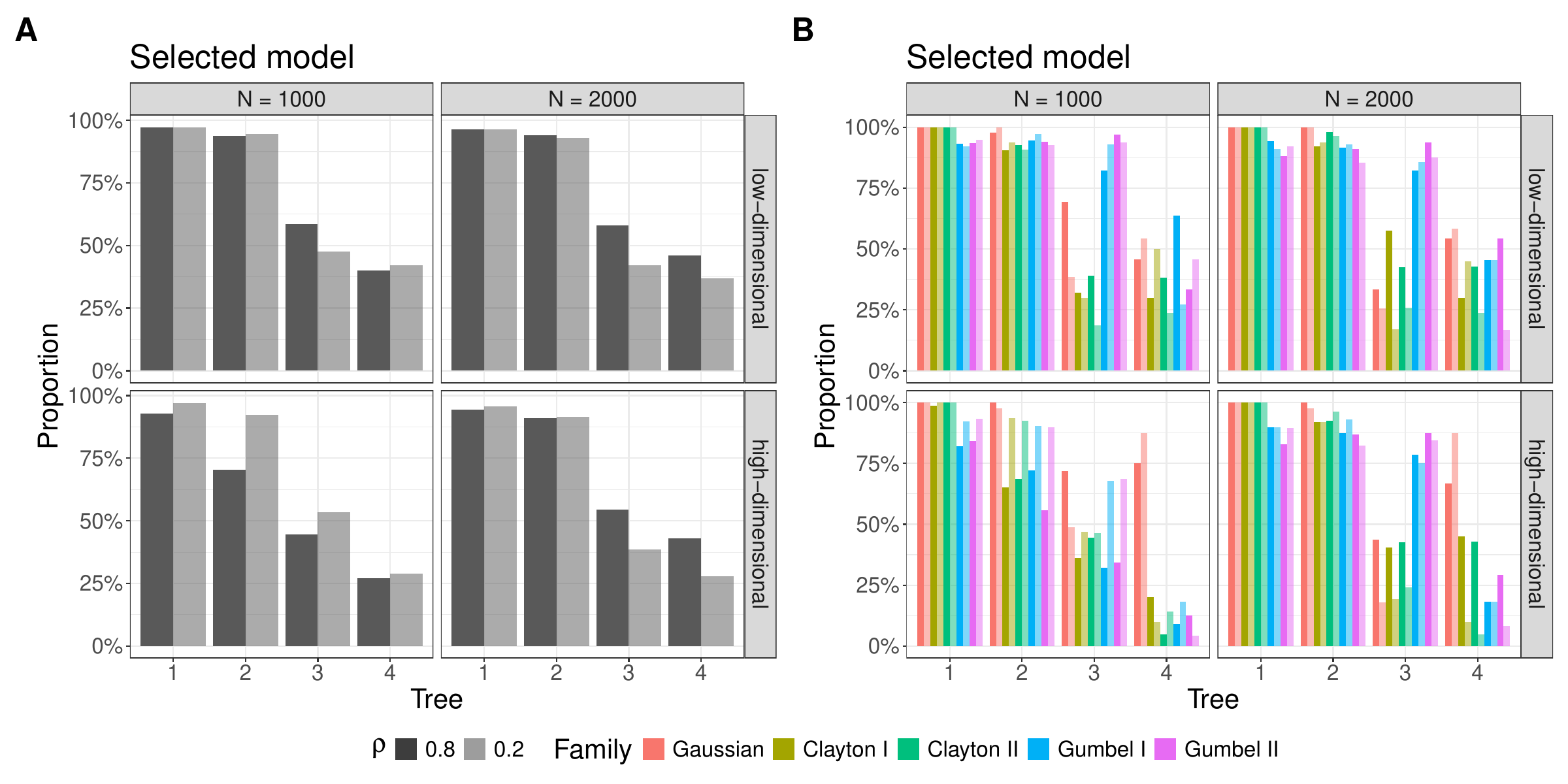}
	\end{center}	
	\caption{Percentage of correctly selected copula families aggregated over all 5 copula families (A) and for a specified copula family (B). The color represent the copula family while the color transparency stands for correlation $\rho$.}
	\label{fig: fams_rvine}
\end{figure}

Figure \ref{fig: fams_rvine} shows the percentage of correctly selected copula families aggregated over all 5 copula families (A) and for a specified copula family (B) at each tree level. Again, we observe in (A) that the estimation errors influence the model fit, as the true copula families are selected mostly correct for the first two tree levels and then the selection rate drops for tree levels 3 and 4. As one might already expect, larger samples sizes tend to guarantee a higher percentage for the selection of the true copula family. However, we can detect no obvious trend in the selection rate across the copula families in (B).

All in all, we conclude that the gradient-boosting including a covariate deselection mechanism (Algorithm \ref{alg: complete gradient-boosting}) selects the informative covariates and estimates its corresponding coefficients quite well for lower tree levels independently of the covariate dimension. Furthermore, the AIC copula family selection criterion selected the true copula family quite accurate in the lower tree levels, with increasing bias in the subsequent trees. In the following section, we additionally demonstrate the benefits of the gradient-boosted conditional vine copulas for temperature forecasting in a low-dimensional covariate setting with 15 covariates.

\section{Application to temperature forecasting}
\label{sec: Application}

Weather forecasting is currently based on numerical weather prediction systems (NWPs), creating an ensemble of weather forecasts, to quantify the forecast uncertainty. As these forecasts suffer from systematic biases and dispersion errors, they may benefit from statistical postprocessing \parencite{Vannitsem2018}.
For postprocessing of univariate weather quantities, various statistical methods have been suggested, as, e.g., ensemble model output statistics (EMOS, \cite{Gneiting2005}). However, statistical independence among the univariate postprocessed marginal distributions of different lead times is sometimes implicitly assumed. This assumption, is often too restrictive and might cause incoherence. Therefore, the corresponding temporal dependence \parencite{Pinson2012b, Lakatos2023} should be restored after the univariate postprocessing to obtain consistent forecasts.

 In the following case study, we focus on restoring temporal dependencies for 2\,\si{\meter} surface temperature forecasts of five different lead times at a single weather station. After a short description of the data set, we present the univariate and multivariate postprocessing methods. We follow the IFM approach, where we first postprocess the temperature forecasts for each lead time separately via EMOS. In a second step, we restore the (temporal) dependence among the lead times by copula approaches to obtain coherent forecasts. Specifically, we compare the Gaussian copula approach \parencite{Moeller2013} and the ensemble copula coupling \parencite{Schefzik2013} with the gradient-boosted conditional vine copulas. Afterwards, we present the results of the copula based approaches in terms of forecast skill and estimated effects.

\subsection{Data}
\label{sec: Data}
The 2\,\si{\meter} surface temperature forecasts obtained from the \textcite{ECMWF2021} consist of 50 perturbed ensemble forecasts for each of the five lead times 24\,\si{\hour}, 48\,\si{\hour}, 72\,\si{\hour}, 96\,\si{\hour}, 120\,\si{\hour}. The forecasts have been initialized at 1200 UTC, have a spatial grid resolution of $25^\circ \times 25^\circ$ ($\approx$ 28\,\si{\km} squared) and are bilinearly interpolated to the spatial coordinates of the city Munich in Germany. The \textcite{DWD2018} provided the 2\,\si{\meter} surface temperature observations for a time range between January 2, 2015 to December 31, 2020, which leads to 2191 observation days over the six years. The observations will be denoted by $Y_j$, where the indices $j\in \mng{1,2,3,4,5}$ refer to the lead times 24\,\si{\hour}, 48\,\si{\hour}, 72\,\si{\hour}, 96\,\si{\hour}, 120\,\si{\hour}, respectively. Additionally, we reduced the $m=50$ member forecast ensemble $\bm{X}_j:=\left(X_{j}^{(1)},\ldots, X_{j}^{(m)}\right)$ to its mean and standard deviation
\begin{align}
\overline{X}_j:=\frac{1}{m}\sum\limits_{i=1}^{m}X_{j}^{(i)},\quad  S_j:=\sqrt{\frac{1}{m-1}\sum\limits_{i=1}^{m}\left(\overline{X}_j-X_{j}^{(i)}\right)^2},
\end{align}
for $j\in \mng{1,2,3,4,5}$, respectively. To take account of seasonal variations, we make use of the forecast initialization day of the year, denoted by $X_{\mathrm{doy}}\in \mng{1,2,\ldots, 366}$. To capture the dependence among the different lead times, we calculated the correlation $\mathrm{Cor}(\bm{X}_i, \bm{X}_j)$ among the forecast ensembles for each lead time combination $i,j\in \mng{1,2,3,4,5}$ with $i<j$. Note, that the realizations of the considered variables will be indicated by lowercase letters. Eventually, we use the period of 2015-2019 as training data set and the remaining days in 2020 as  validation data set. The data is part of the data set by \textcite{Jobst2023e}.

\subsection{Methods}

As already indicated, we follow the IFM approach for the multivariate postprocessing of 2\,\si{\meter} surface temperature of different lead times. Therefore, we firstly apply a univariate postprocessing method for each lead time separately and afterwards we restore the temporal dependencies via copula approaches. 

\subsubsection{Univariate Postprocessing}

For the univariate postprocessing of 2\,\si{\meter} surface temperature forecasts, we make use of the ensemble model output statics (EMOS) originally suggested by \textcite{Gneiting2005}. Specifically, we assume a normal distribution $Y_j\sim\mathcal{N}(\mu_j, \sigma_j^2)$ for the predictive distribution of each lead time case $j\in \mng{1,2,3,4,5}$. Following \textcites{Lang2020, Jobst2024}, we employ a smooth EMOS version, i.e. we assume 
\begin{align}
	\mu_j:=a_{0,j} + f_{0,j} + (a_{1,j} + f_{1,j})\cdot \overline{x}_j,\quad \log(\sigma_j):=b_{0,j} + g_{0,j} + (b_{1,j} + g_{1,j})\cdot s_j,
\end{align}
with coefficients $a_{0,j},a_{1,j},b_{0,j},b_{1,j}\in \R$, ensemble mean $\overline{x}_j$ and standard deviation $s_j$ for $j\in \mng{1,2,3,4,5}$. Furthermore, 
{\footnotesize
\begin{align}
	f_{k,j}&:=\alpha_{k,j,1}\sin\left(\frac{2\pi x_{\mathrm{doy}}}{365.25}\right)+\alpha_{k,j,2}\cos\left(\frac{2\pi x_{\mathrm{doy}}}{365.25}\right)+\alpha_{k,j,3}\sin\left(\frac{4\pi x_{\mathrm{doy}}}{365.25}\right)+\alpha_{k,j,4}\cos\left(\frac{4\pi x_{\mathrm{doy}}}{365.25}\right),\\
	g_{k,j}&:=\beta_{k,j,1}\sin\left(\frac{2\pi x_{\mathrm{doy}}}{365.25}\right)+\beta_{k,j,2}\cos\left(\frac{2\pi x_{\mathrm{doy}}}{365.25}\right)+\beta_{k,j,3}\sin\left(\frac{4\pi x_{\mathrm{doy}}}{365.25}\right)+\beta_{k,j,4}\cos\left(\frac{4\pi x_{\mathrm{doy}}}{365.25}\right),
\end{align}
}%
denote truncated Fourier series to model the cyclic seasonal behavior in the intercept and slope with coefficients  $\alpha_{k,j,l}, \beta_{k,j,l}\in \R$ for $k=0,1,\, j=1,2,3,4,5,\, l=1,2,3,4$. All coefficients are optimized via the continuous ranked probability score (CRPS, \textcite{Matheson1976}) with the training data set as static training period for each lead time separately. For the model estimation we make use of the \texttt{R}-package \texttt{crch} by \textcite{Messner2016}.

\subsubsection{Multivariate postprocessing} 

In a second step we re-obtain the temporal dependencies among the 5 different lead times via different copula approaches. For that, we denote the postprocessed marginal distribution functions by $F_1,\ldots, F_5$ for the 5 different lead times and consider the temporally ordered random vector $F\sim (Y_1,\ldots, Y_5)$. In the following, we briefly summarize two common multivariate postprocessing techniques and present the setting for the conditional vine copula approach.

\paragraph{Ensemble copula coupling.}

The ensemble copula coupling (ECC, \cite{Schefzik2013}) is based on the assumption, that the raw ensemble forecasts can capture the true multivariate dependence structure. Given the postprocessed univariate marginal distributions $F_1, \ldots, F_5$, the method is based on two steps:
\begin{enumerate}
	\item Draw a sample $\widehat{x}_j^{(1)}, \ldots, \widehat{x}_j^{(m)}$ of the same size as the raw ensemble from each postprocessed distribution $F_j$ for $j=1,\ldots, 5$, which will be arranged in ascending order.
	\item Calculate the permutations $\bm{\pi}_j:=(\pi_j(1),\ldots, \pi_j(m))$ by the rank order structure of the raw ensemble $\pi_j(k):=\text{rank}(\widehat{x}_j^{(k)})$ for $k=1,\ldots, m$ with ties resolved at random for each postprocessed distribution $F_j$. To obtain the ECC-corrected forecast ensemble, the sample from step 1 is reordered according to the permutation $\bm{\pi}_j$ via
\begin{align}
	\tilde{x}_j^{(k)}:=\widehat{x}_{\pi_{j}(k)}^{(k)},\quad k=1,\ldots,m,
\end{align}
for each postprocessed distribution $F_j$ for $j=1,\ldots,5$.
\end{enumerate}
Different procedures for the drawing the sample in step 1 have been considered \parencites{Schefzik2013, Hu2016}. To ensure a fair comparison among the methods, we draw $m$ random samples 
\begin{align}
   \widehat{x}_j^{(1)}:=F_j^{-1}(u_1), \ldots, \widehat{x}_j^{(m)}:=F_j^{-1}(u_m),
\end{align}
from independently, uniform distributed samples $u_1,\ldots, u_m$ on  (0,1) for $j=1,\ldots, 5$. The method is implemented by an own \texttt{R}-code. 

\paragraph{Gaussian copula approach.}
For the Gaussian copula approach (GCA, \cite{Moeller2013}), we assume that the dependence among the probability integral transformed observations $u_j=F_j(y_j)$ for $j=1,\ldots, 5$ can be captured by a multivariate Gaussian copula, i.e. 
\begin{align}
	C(u_1,\ldots, u_5)=\Phi_5(\Phi^{-1}(u_1), \ldots, \Phi^{-1}(u_5)\,\vert\, \bm{\Sigma}),
\end{align}
where $\Phi_5(\cdot \,\vert\, \Sigma)$ denotes the cumulative distribution function of a $5$-dimensional normal distribution with mean zero, correlation matrix $\bm{\Sigma}$, and $\Phi^{-1}$ denoting the quantile function of the univariate standard normal distribution $\Phi$. The GCA consists of the following four steps:

\begin{enumerate}
	\item Latent standard Gaussian observations are calculated via 
\begin{align}
\tilde{y}_j:=\Phi^{-1}(F_j(y_j)),	
\end{align}
employing a set of past observations (here all available observations in the training data set) and the postprocessed marginal distributions $F_j$ for all $j=1,\ldots, 5$.
	\item A parametric (here empirical) $5\times 5$ correlation matrix $\widehat{\bm{\Sigma}}$ of the $5$-dimensional normal distribution $\Phi_5$ is estimated from the latent standard Gaussian observations in step 1. 
	\item Multivariate random samples $\bm{Z}_1,\ldots, \bm{Z}_m$  are drawn from $\mathcal{N}_5(\bm{0},\widehat{\bm{\Sigma}})$. 
	\item In a last step, the GCA postprocessed forecast ensemble of the validation data set is calculated via
\begin{align}
	\tilde{x}_j^{(k)}:=F_j^{-1}(\Phi(z_j^{(k)})),\quad k=1,\ldots, m,
\end{align}
of each postprocessed univariate distribution $F_j$ for $j=1,\ldots, 5$.
\end{enumerate}
Note, that combining a multivariate Gaussian copula with Gaussian marginal distributions is equal to a multivariate Gaussian distribution. The implementation of this method is based on the function  \texttt{rmvnorm} of the \texttt{R}-package \texttt{Rfast} by \textcite{Papadakis2021}.

\paragraph{Gradient-boosted conditional vine copula.}
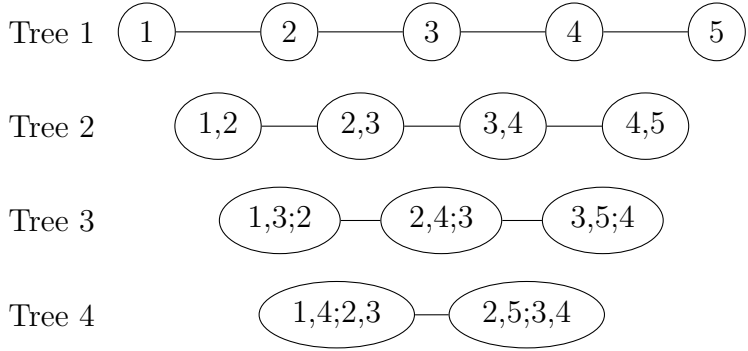
\begin{figure}[h!]
\begin{center}
\begin{tikzpicture}[scale = 1.25]
\usetikzlibrary{shapes}
	\node (Tree 1) at (-1, 0) {Tree 1};
	\node (Tree 2) at (-1, -1) {Tree 2};
	\node (Tree 3) at (-1, -2) {Tree 3};
	\node (Tree 4) at (-1, -3) {Tree 4};

      \node[ellipse, minimum size=0.75cm, draw=black] (1) at (0,0) {1};
      \node[ellipse, minimum size=0.75cm, draw=black] (2) at (1.5, 0) {2};
      \node[ellipse, minimum size=0.75cm, draw=black] (3) at (3, 0) {3};
      \node[ellipse, minimum size=0.75cm, draw=black] (4) at (4.5,0) {4};
            \node[ellipse, minimum size=0.75cm, draw=black] (5) at (6,0) {5};
			
\draw (1)--node[midway,above,sloped]{}(2)
			(2)--node[midway,above,sloped]{}(3)
			(3)--node[midway,above,sloped]{}(4)
			(4)--node[midway,above,sloped]{}(5);					

      \node[ellipse, minimum size=0.75cm, draw=black] (1) at (0.75, -1) {1,2};
      \node[ellipse, minimum size=0.75cm, draw=black] (2) at (2.25, -1) {2,3};
      \node[ellipse, minimum size=0.75cm, draw=black] (3) at (3.75, -1) {3,4};
      \node[ellipse, minimum size=0.75cm, draw=black] (4) at (5.25, -1) {4,5};
			
\draw (1)--node[midway,above,sloped]{}(2)
			(2)--node[midway,above,sloped]{}(3)
			(3)--node[midway,above,sloped]{}(4);	

      \node[ellipse, minimum size=0.75cm, draw=black] (1) at (1.4,-2) {1,3;2};
      \node[ellipse, minimum size=0.75cm, draw=black] (2) at (3.1, -2) {2,4;3};
      \node[ellipse, minimum size=0.75cm, draw=black] (3) at (4.8, -2) {3,5;4};
			
\draw (1)--node[midway,above,sloped]{}(2)
			(2)--node[midway,above,sloped]{}(3);			
			
      \node[ellipse, minimum size=0.75cm, draw=black] (1) at (2.0,-3) {1,4;2,3};
      \node[ellipse, minimum size=0.75cm, draw=black] (2) at (4, -3) {2,5;3,4};
			
\draw (1)--node[midway,above,sloped]{}(2);				
			
\end{tikzpicture}	
\end{center}
	\caption{5-dimensional D-vine.}\label{fig: D-vine}
\end{figure}
For setting up a conditional vine copula, we first need to specify a regular vine. We choose a subclass, of a regular vine, called drawable vine (D-vine), where each node in each tree of the regular vine is connected at most to two other nodes (see Figure \ref{fig: D-vine}). As our data consists of a natural sequential ordering in the lead times $1-2-3-4-5$, a D-vine is particularly suited for this scenario (see, e.g. \cites{Smith2010, NaiRuscone2016, Czado2019}). For all conditional bivariate copulas we specify a GLM with the 15 covariates
$$\bm{Z}=\left(1,\quad \left(\sin\left(\frac{2\pi k X_{\mathrm{doy}}}{365.25}\right),\cos\left(\frac{2\pi k X_{\mathrm{doy}}}{365.25}\right)\right)_{k=1,2},\quad (\mathrm{Cor}(\bm{X}_i,\bm{X}_j))_{1\leq i<j \leq 5}\right).$$
A seasonal varying intercept can be captured by the sine- and cosine transformed day of the year, while the correlations among the forecast ensembles of pair-wise different lead times are used to integrate information about the inter-temporal dependence. We estimate the conditional D-vine copula via gradient-boosting and covariate deselection (Algorithm \ref{alg: complete gradient-boosting}). A maximum number of boosting iterations $m_{\text{stop}}=500$, step length $\nu=0.1$ and deselection threshold $\gamma = 0.01$ as in Section \ref{sec: Sim conditional vine copula} turned out to be suited on the training data set. Furthermore, we specify two conditional vine copula models: one model (CVC) for which we allow all five copula families from the simulation study for each conditional bivariate copula and another model (CVC-G), where we select the Gaussian copula family for each conditional bivariate copula. In the latter case, the conditional D-vine copula  corresponds to a conditional multivariate Gaussian copula \parencite{Czado2019} in contrast to GCA, where we employ an unconditional multivariate Gaussian copula. Note, that the empirical correlation matrix of the observations shows with increasing lead time a reduction in the correlations, which reminds of a Toeplitz structure similar as employed in our simulation studies.

\subsection{Verification methods}

As we focus in this application on the verification of multivariate predictive distributions $F$, we make use of the \textit{energy score} (ES, \cite{Gneiting2007}), which can be seen as an multivariate extension of the CRPS. Given a $d$-dimensional multivariate forecast $F$ in terms of an $m$-member forecast ensemble $\bm{x}^{(1)},\ldots, \bm{x}^{(m)}\in \R^d$  and corresponding observation $\bm{y}=(y_1,\ldots,y_d)\in\R^d$, the energy score is obtained by 
\begin{align}
\text{ES}(F,\bm{y})=\frac{1}{m}\sum\limits_{k=1}^{m}\norm{\bm{x}^{(k)}-\bm{y}}-\frac{1}{2m^2}\sum\limits_{k=1}^{m}\sum\limits_{j=1}^{m}\norm{\bm{x}^{(k)}-\bm{x}^{(j)}},
\end{align}
where $\norm{\,\cdot\,}$ denotes the Euclidean norm on $\R^d$. Furthermore, if $F$ is given as multivariate forecast distribution function, the energy score can be obtain by the Monte-Carlo approximation
\begin{align}
\text{ES}(F,\bm{y})=\frac{1}{m}\sum\limits_{k=1}^{m}\norm{\bm{x}^{(k)}-\bm{y}}-\frac{1}{2(m-1)}\sum\limits_{k=1}^{m-1}\norm{\bm{x}^{(k)}-\bm{x}^{(k+1)}},
\end{align}
where $\bm{x}^{(1)}, \ldots, \bm{x}^{(m)}$ are samples issued from $F$ \parencite{Gneiting2008}.
As it is advised to use multiple scores for assessing the predictive performance we additionally consider the \textit{variogram score} of order $p$ ($\text{VS}^p$, \cite{Scheuerer2015b}) as a multivariate proper scoring rule. The score is defined as
\begin{align}\text{VS}^p(F,\bm{y})=\sum\limits_{i=1}^{d}\sum\limits_{j=1}^{d}\omega_{ij}\left(\abs{y_i-y_j}-\frac{1}{m}\sum\limits_{k=1}^{m}\abs{x_{i}^{(k)}-x_{j}^{(k)}}^p\right)^2,
\end{align}
where $\omega_{ij}\geq 0$ is the weight for pair $(i,j)$. Compared to other multivariate proper scoring rules, such as e.g. the energy score, the $\text{VS}^p$ is more sensible to correlation misspecifications. We apply an unweighted version of the $\text{VS}^p$, i.e. $\omega_{ij}=1$ for all pairs $(i,j)$ with order $p=0.5$ following the suggestions of \textcite{Scheuerer2015b}. For the calculation of both scores, we choose a sample size of $m=10000$ for a fair comparison among GCA, CVC and CVC-G, while we are by construction restricted to a sample size of $m=50$ for the raw ensemble and ECC. Eventually, to assess the statistical improvements in terms of the considered scores, we conduct a Diebold-Mariano test \parencite{Diebold1995} with automatic lag selection according to \textcite{Newey1994} and Bartlett weights for the heteroscedasticity and autocorrelation consistent variance estimator. We employ a significance level of $\alpha = 0.05$ for the following tests. 




\subsection{Results}

\begin{figure}[h!]
	\begin{center}
		\includegraphics[scale = 0.3]{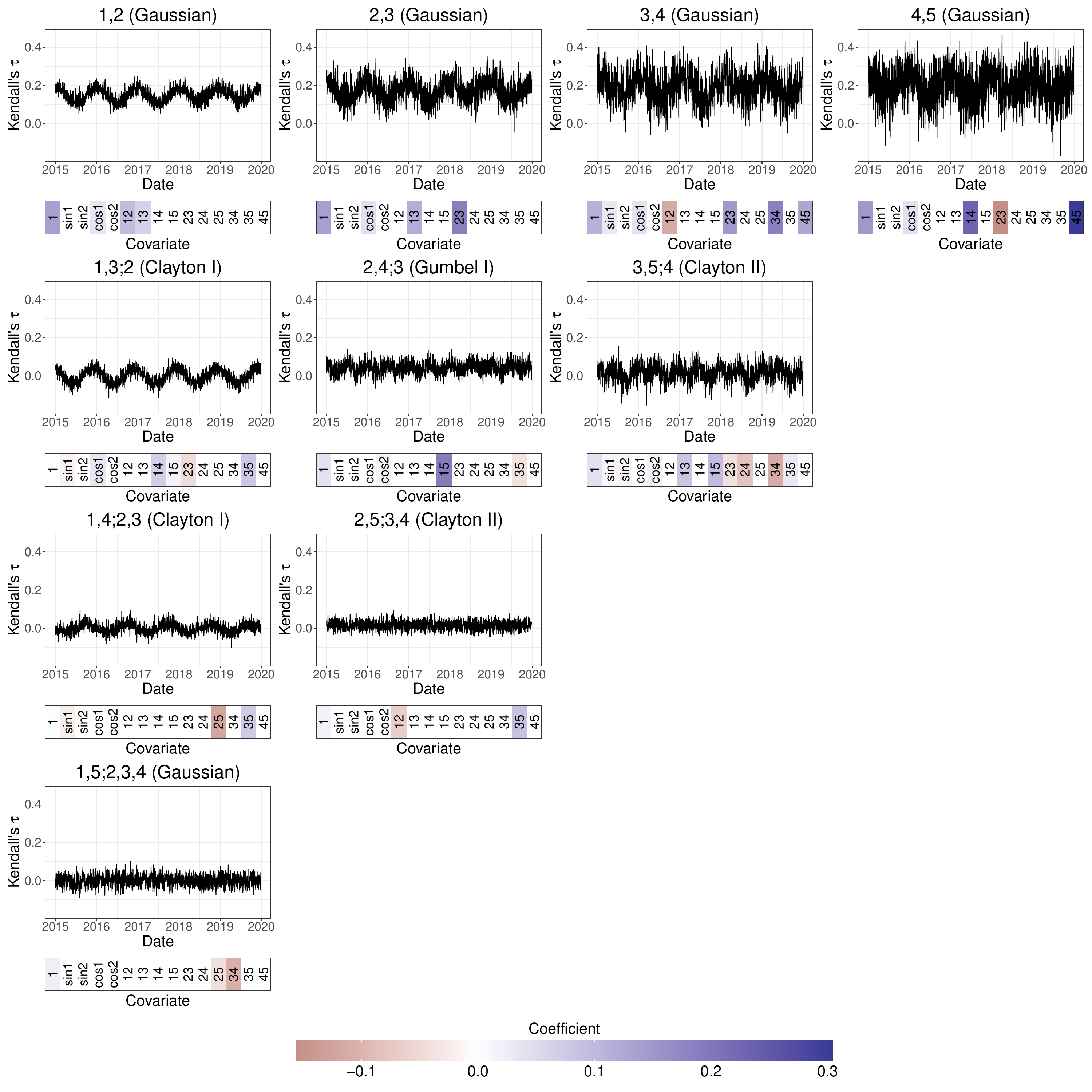}
	\end{center}	
	\caption{Kendall's $\tau$ correlation time series for each conditional bivariate copula (family) from tree level 1 (first row) to tree level 4 (last row) for CVC in the training data set. The corresponding color gradient indicates the value of the coefficient of the (selected) covariates for each conditional bivariate copula.}
	\label{fig: kendall_coef_app}
\end{figure}

Table \ref{tab: scores} shows the mean scores of the raw ensemble as well as the ECC-, GCA-, CVC- and CVC-G-postprocessed ensemble forecasts in the validation data set. All postprocessing methods are able to considerably improve the raw ensemble with respect to $\text{VS}^{0.5}$ and ES. Furthermore, CVC-G and CVC yield to significant improvements in terms of the considered scores over ECC and GCA. However, there are minor score differences among CVC-G and CVC suggesting, that the Gaussian dependence assumption is sufficient. The reduced forecast skill of CVC in comparison to CVC-G might be traced back to the slight overfitting in terms of the copula family selection (see also Figure \ref{fig: kendall_coef_app}). As we can already conclude from the $\text{VS}^{0.5}$ of CVC-G and CVC, conditionally modelling the temporal dependencies among the lead times can yield an improvement over models with constant dependence assumption such as GCA.

\begin{table}[h!]
\begin{center}
\begin{tabular}{c c c c c c} 
\toprule
 & Raw ensemble & ECC & GCA & CVC-G & CVC \\ \hline
$\text{VS}^{0.5}$ & $6.778^{\,(0.055,\, \underline{0.058})}$ & $6.695^{\,(0.015,\, \underline{0.016})}$ & $6.632^{\,(0.002,\, \underline{0.002})}$ & \textbf{6.587}$^{\,(\underline{0.876})}$ & $6.589^{\,(0.124)}$ \\ 
ES & $3.010^{\,(0.000,\, \underline{0.000})}$ & $2.784^{\,(0.000,\, \underline{0.000})}$ & $2.747^{\,(0.040,\, \underline{0.045})}$ & \textbf{2.744}$^{\,(\underline{0.667})}$ & $2.744^{\,(0.333)}$ \\
\bottomrule
\end{tabular}
\end{center}
\caption{Verification scores aggregated over all time points in the validation data set. Bold values represent the best value for each score. The score exponents show the $p$-values in favor of the CVC-G (not underlined) and CVC (underlined) over the competing methods according to the Diebold-Mariano test.}
\label{tab: scores}
\end{table}

\begin{figure}[h!]
	\begin{center}
		\includegraphics[scale = 0.4]{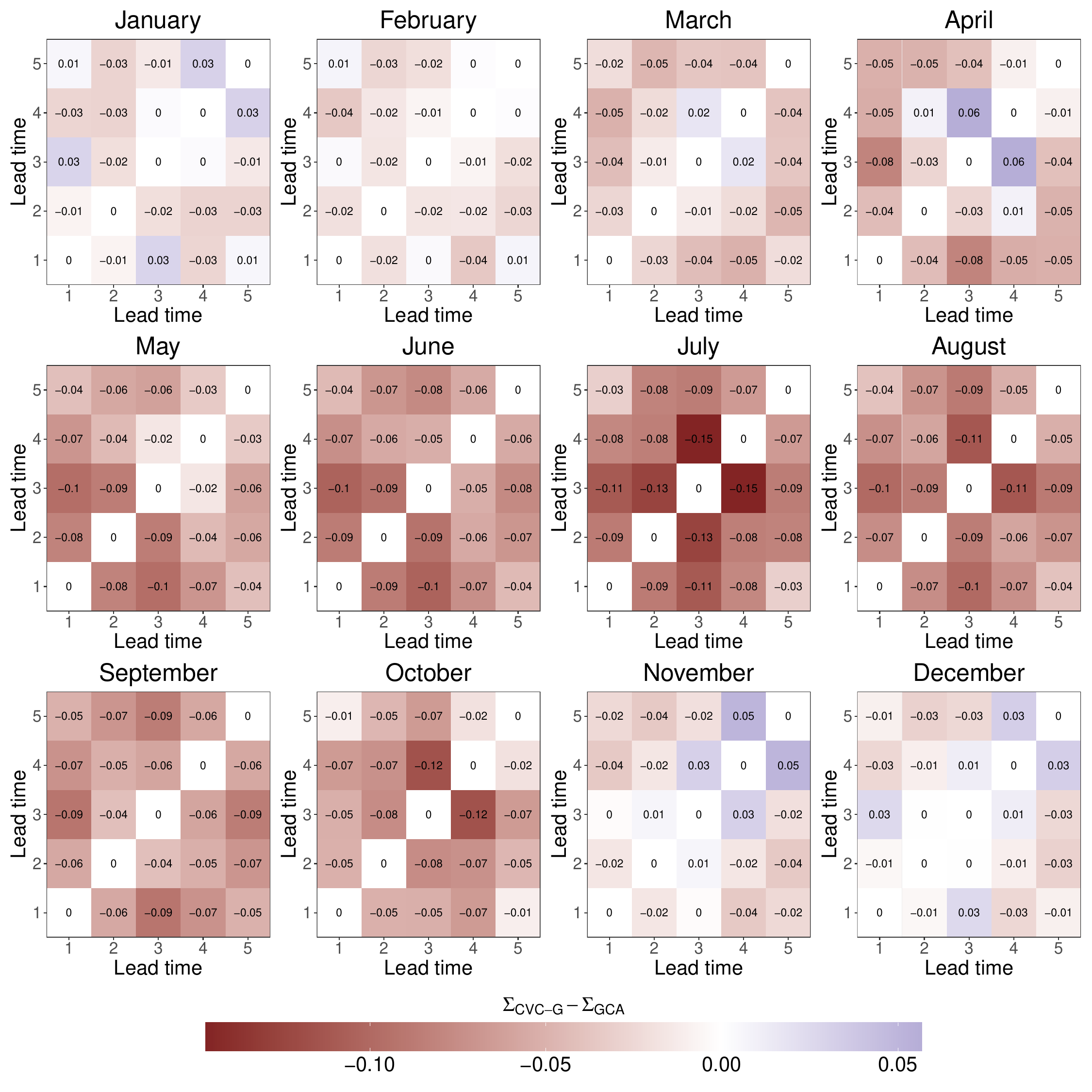}
	\end{center}	
	\caption{Monthly averaged differences between the correlation matrices of CVC-G and GCA in the validation data set.}
	\label{fig: cor_plots}
\end{figure}

In Figure \ref{fig: cor_plots}, the monthly averaged differences  between the correlation matrices of CVC-G and GCA are visualized. Roughly speaking, we observe clearly lower correlations (dark red) among the different lead time forecasts for CVC-G than for GCA in the summer and fall months. In contrast, there are overall smaller correlation differences in winter months among CVC-G and GCA. However, CVC-G yields for the same period also slightly higher correlations (light purple) among the forecasts of higher lead times than this is the case for GCA. These results indicate, that the correlation among the lead time forecasts changes over time and shows seasonal dependencies.

Figure \ref{fig: kendall_coef_app} shows the predicted Kendall's $\tau$ correlation time series for each conditional bivariate copula depending on the tree level in the training data set. First of all, we can confirm the already detected seasonal patterns for Kendall's $\tau$ correlation induced by the seasonal intercept in the GLM. Interestingly, we observe higher correlations among the lead times in winter than in summer. This might be related to the fact, that the marginal distributions of all lead times show higher variances in winter and lower variances in summer as well lower mean in winter than in summer. Furthermore, forecasts with higher lead times exhibit higher Kendall's $\tau$ correlations amongst each other, and a higher variance in Kendall's $\tau$ correlations than forecasts with lower lead times. This is specifically pronounced in the first tree. Additionally, the seasonal effect in the Kendall's $\tau$ correlation time series and consequently its volatility reduces from tree level 1 to tree level 4. This is in line with our previous findings, that for the higher tree levels constant copula parameters might be sufficient. As the Kendall's $\tau$ correlations fluctuate mostly around zero for tree levels greater 2, truncating the conditional D-vine copula at tree level 2, i.e. assuming Independence copulas for all bivariate copulas after tree level 2 is a possible course of action. This approach yields only to small changes for CVC-G and CVC in terms of the considered scores (results not shown here).

The grids below each Kendall's $\tau$ time series in Figure \ref{fig: kendall_coef_app} show the color-coded estimated coefficient of the covariates for each conditional bivariate copula in CVC. Covariate \texttt{1} represents the intercept and covariates \texttt{sin1}, \texttt{cos1}, \texttt{sin2}, \texttt{cos2} correspond to the sine- and cosine transformed day of the year for order $k=1,2$, respectively. Additionally, e.g. covariate \texttt{25} corresponds to the correlation among the forecast ensemble $\bm{X}_2$ and $\bm{X}_5$, i.e. $\text{Cor}(\bm{X}_{2}, \bm{X}_5)$. We observe that non-zero coefficients appear most frequently for tree level 1 and 2, which indicates that the dependence among the lead times can be influenced by covariates. As one may expect, copula 1,2 selects the corresponding lead time correlation-based covariate \texttt{12} in the first tree. Analogously, copula 2,3, copula 3,4 and copula 4,5 select its corresponding lead time correlation-based covariate \texttt{23}, \texttt{34}, \texttt{45}, respectively. Additionally, its corresponding coefficient has a high value in contrast to the other coefficients, underlying its importance as predictor. In the second tree level, we can see that e.g. copula 3,5;4 selects the lead-time related covariate \texttt{35} but also covariates \texttt{15} and \texttt{24} which might not be expected as predictors. Similar observations can be made for the other two copulas in the second tree level. 

All in all, we conclude that CVC-G and CVC are able to significantly outperform state-of-the-art multivariate ensemble postprocessing methods and that a seasonally adaptive model for the temporal dependence among the lead time forecasts can considerably improve the predictive performance. However, explicitly linking the lead-time correlation-based covariates to the corresponding conditional bivariate copulas should be carefully considered. 


\section{Conclusion and outlook}
\label{sec: Conclusion and outlook}

In this work we introduced conditional vine copulas within a gradient-boosting estimation framework. Covariates are linked in terms of a GLM to the Kendall's $\tau$ correlation coefficient, from which the parameters for conditional bivariate copulas can be derived. The gradient-boosting Algorithm \ref{alg: complete gradient-boosting} allows a natural covariate selection and coefficient estimation adapted for sparse low- and high-dimensional covariate settings. Additionally, we extended the estimation procedure to conditional vine copulas. 

In a simulation study we investigated a pre-specified GLM for 5 different bivariate copula families and for 5-dimensional conditional vine copulas in low- as well as high dimensional covariate settings. The results for the conditional bivariate copula fitting showed, that Algorithm \ref{alg: complete gradient-boosting} is able to detect the true effects and to select mostly the informative covariates in nearly each simulation scenario. Furthermore, AIC outperformed CV as stopping criterion and the AIC is suitable to identify the (true) copula family. For the estimation of the conditional vine copulas we observe, that the estimated effects are almost unbiased in the first two tree levels and that the (de)selection of (non-)informative covariates works well. Furthermore, increasing the sample size can provide more accurate fits. But there is an increasing shrinkage towards zero in the subsequent trees, which can yield to a biased copula family and covariate selection. However, this is a general issue for the estimation of conditional vine copulas already indicated by \textcite{Vatter2018} and not only related to gradient-boosting. 

In a case study for the lead time based postprocessing of 2\,\si{\meter} surface temperature forecasts, CVC and CVC-G outperform benchmark methods in terms of variogram and energy score, indicating that the use of carefully selected covariates to predict the temporal dependencies can improve the multivariate performance. Besides of a promising forecast skill, CVC and CVC-G provided interesting insights in the covariate selection for the conditional bivariate copulas.

Despite of its merits, we detect some limitations of the conditional vine copulas in its current formulation. Firstly, our approach is not yet suited for discrete or mixed continuous-discrete responses. However, with slight modifications for the parameterization of the copula families, the gradient-boosting based estimation of conditional vine copulas can be extended based on the existing results for discrete \parencite{Panagiotelis2012} or mixed continuous-discrete \parencite{Stoeber2013b} responses.
On top of our agenda is an extension of the conditional vine copulas in a gradient-boosting framework to the class of structured additive regression models (STAR, \cite{Fahrmeir2021}).
To reduce the increasing bias for the covariate effects in higher tree levels, we could first apply Algorithm \ref{alg: complete gradient-boosting} for the covariate selection. In a second step, the conditional bivariate copulas could be fitted on the reduced covariate set using the (penalized) maximum-likelihood estimation procedure by \textcite{Vatter2018}. 
Moreover, we only analyzed one-parametric copula families. However, the use of two-parametric copula families, such as e.g. Student-$t$ copula where the second parameter depends on covariates as well would allow an even more flexible dependence modeling. To overcome this shortcoming, the work of \textcite{Thomas2017} using a non-cyclic parameter updating with an adaptive step length \parencite{Zhang2022} in the gradient-boosting algorithm could be applied. Furthermore, in terms of non-sparse covariate scenarios, the cumulative risk suggested by \textcite{Stroemer2021} could be used in the \texttt{R}-package \texttt{boostCopula} to avoid a deselection of too many covariates.
With regard to applications, it would be interesting to conduct a more extensive comparison of our suggested method with other multivariate models in the field of ensemble postprocessing as well as for other applications.

\section*{Acknowledgements}
We are grateful to the European Centre for Medium-Range Weather Forecasts (ECMWF) and the German Weather Service (DWD) for providing forecasts and observation data, respectively. Furthermore, the authors acknowledge support of the research by Deutsche Forschungsgemeinschaft (DFG) Grant Number 395388010. Annette M\"oller acknowledges support by Deutsche Forschungsgemeinschaft (DFG) Grant Number 520017589.

\clearpage
\pagebreak
\newpage
\appendix
\begin{Large}
\textbf{Appendix}	
\end{Large}
\section{Conditional bivariate copulas}
\label{app: Conditional bivariate copulas}

\begin{figure}[h!]
	\begin{center}
		\includegraphics[scale = 0.4]{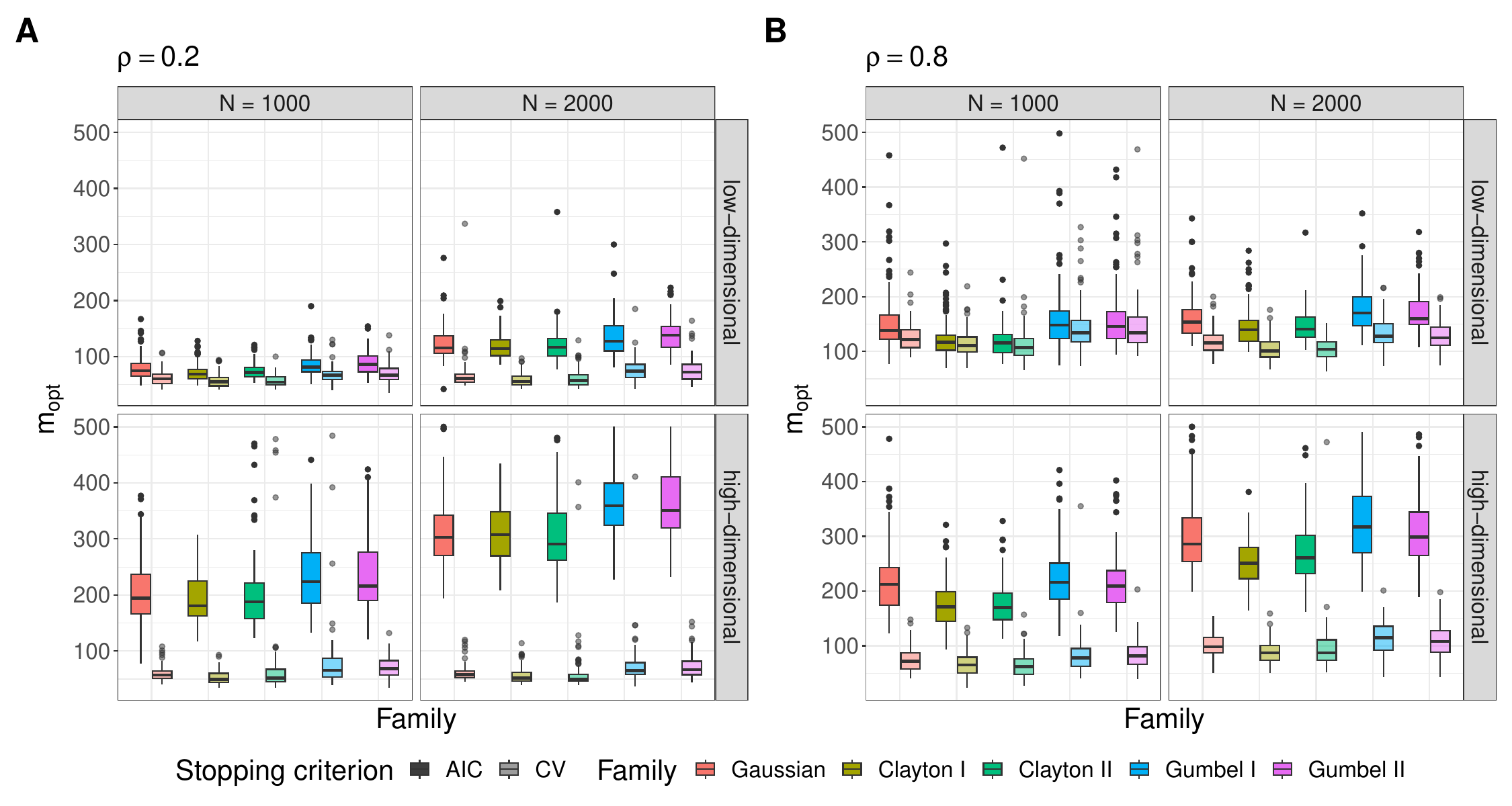}
	\end{center}	
	\caption{Optimal number of boosting iterations $m_{\text{opt}}$ determined by AIC or CV for the low-correlated (A) and high-correlated (B) setting. The color represent the copula families while the color transparency stands for the stopping criterion.}
	\label{fig: mopt_bicop_aic}
\end{figure}

\begin{figure}[h!]
	\begin{center}
		\includegraphics[scale = 0.3]{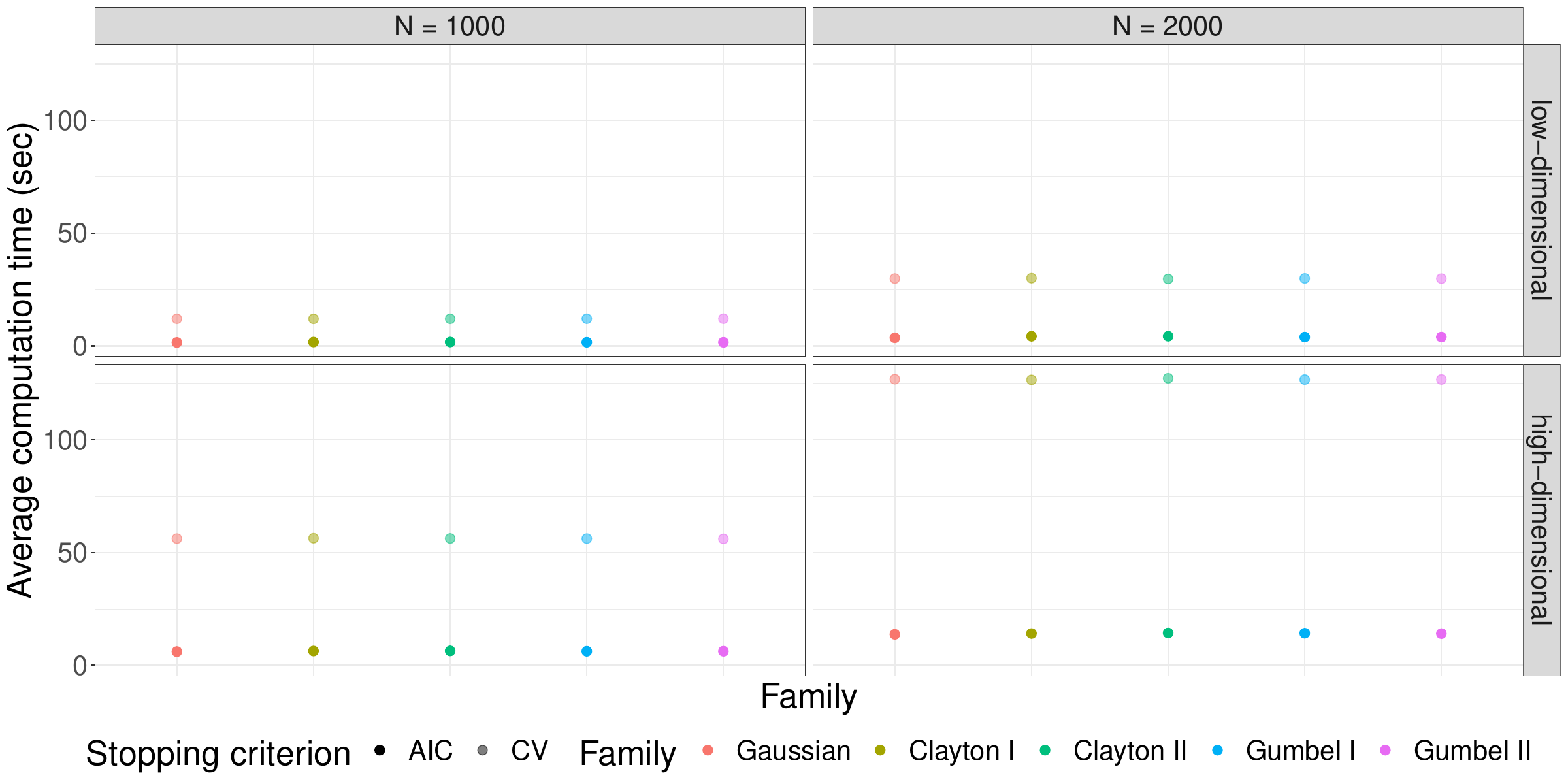}
	\end{center}	
	\caption{Computation times averaged over the the correlation settings $\rho\in\mng{0.2, 0.8}$. The color represents the copula families while the color transparency stands for the stopping criterion.}
	\label{fig: time_bicop_aic}
\end{figure}

\begin{figure}[h!]
	\begin{center}
		\includegraphics[scale = 0.4]{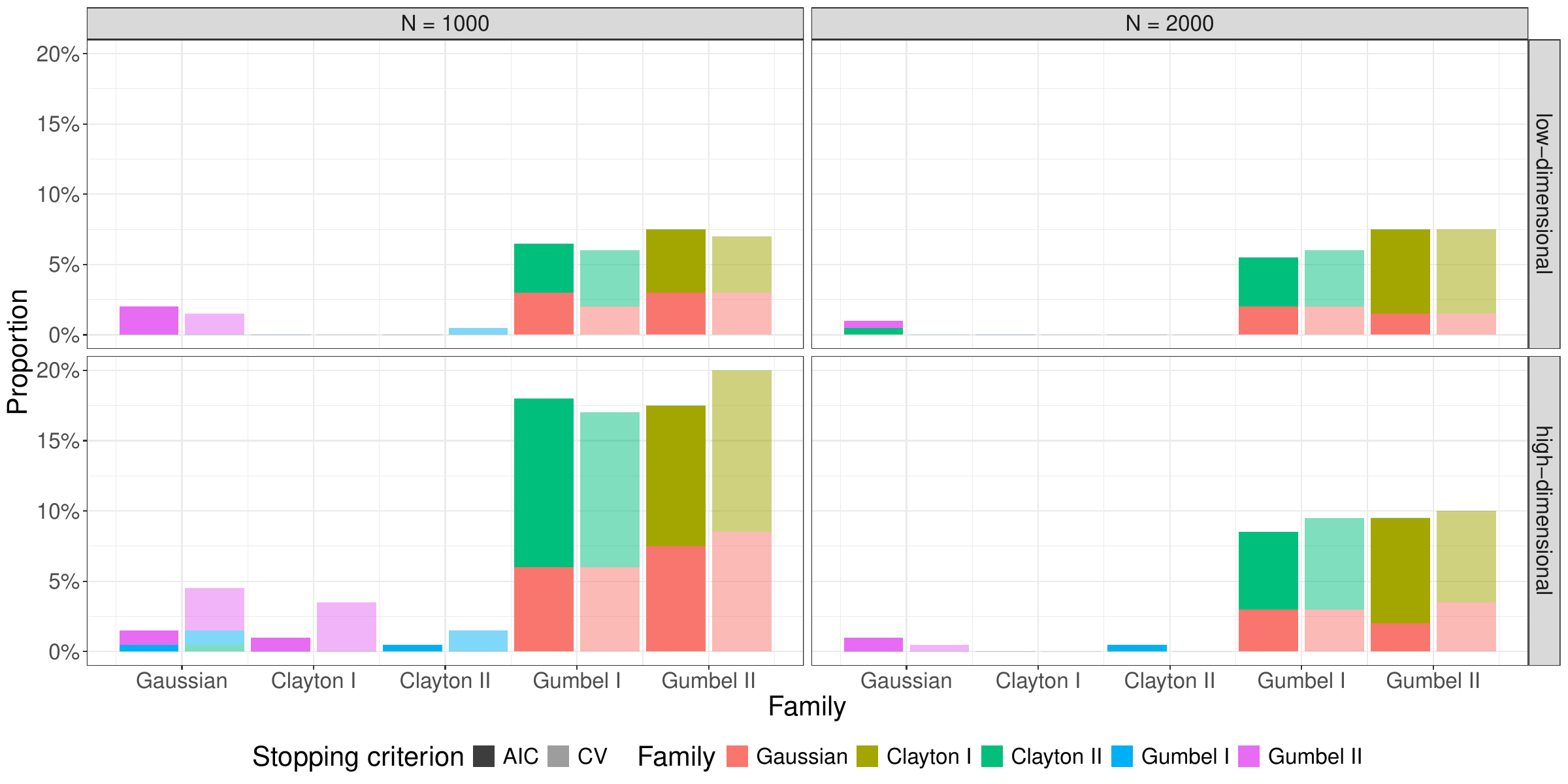}
	\end{center}	
	\caption{Percentage of false selected copula families for each specified copula family based on (in-sample) log-likelihood aggregated over the low-correlated and high-correlated setting. The color represents the copula families while the color transparency stands for the stopping criterion.}
	\label{fig: fams_bicop_loglik}
\end{figure}

\begin{figure}[h!]
	\begin{center}
		\includegraphics[scale = 0.4]{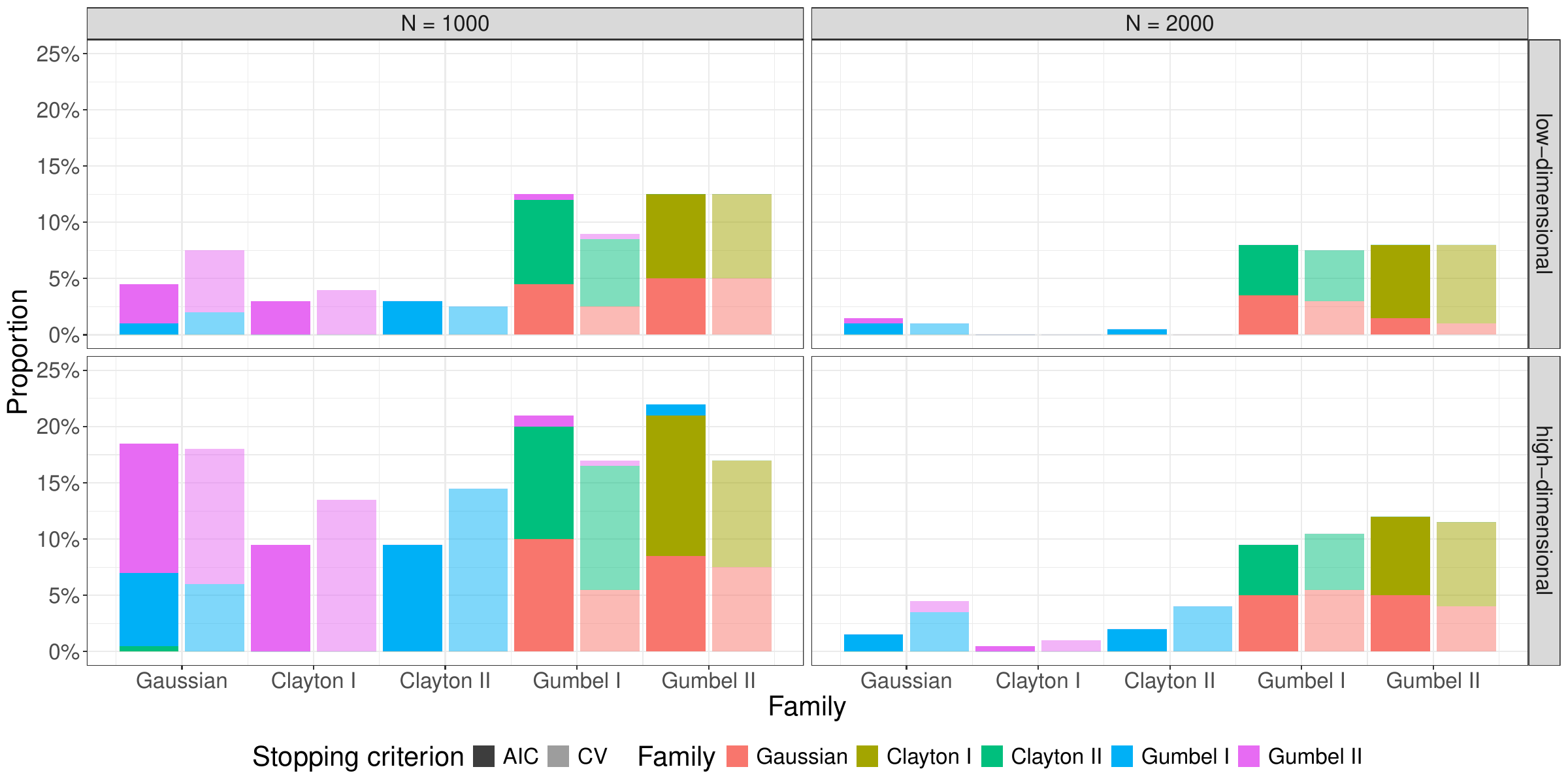}
	\end{center}	
	\caption{Percentage of false selected copula families for each specified copula family based on predictive risk using 25\% of the training data for out-of-sample predictions aggregated over the low-correlated and high-correlated setting. The color represent the copula families while the color transparency stands for the stopping criterion.}
	\label{fig: fams_bicop_pr}
\end{figure}

\newpage
\pagebreak
\clearpage
\section{Conditional vine copulas}
\label{app: Conditional vine copulas}

\begin{figure}[h!]
	\begin{center}
		\includegraphics[scale = 0.4]{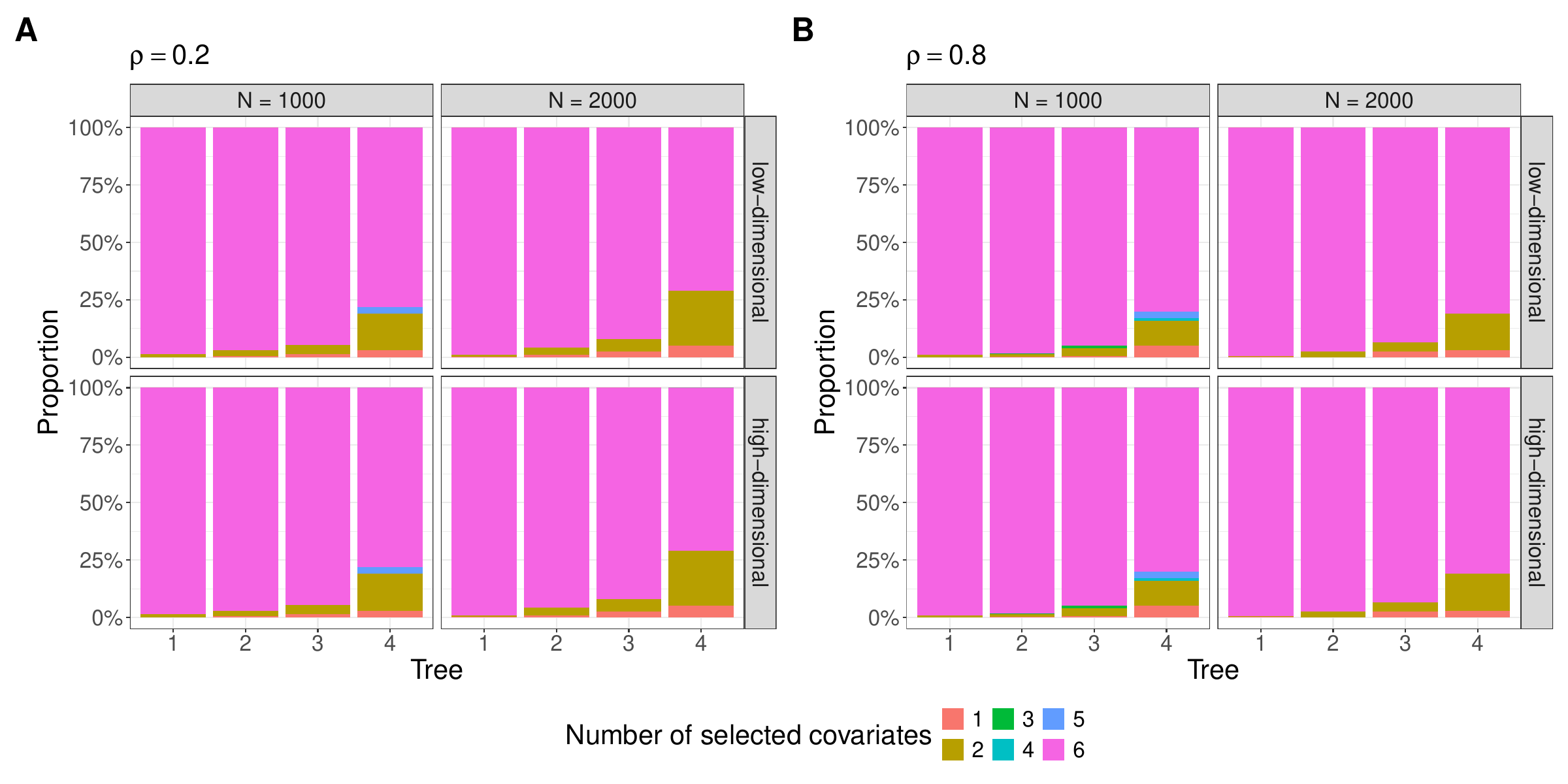}
	\end{center}	
	\caption{Bar charts for the proportion (\%) of selected covariates across all copulas for each tree level and $n=100$ repetitions in the ``Specified'' model for the low-correlated (A) and high-correlated (B) setting.}
	\label{fig: tp_rvine}
\end{figure}

\begin{figure}[h!]
	\begin{center}
		\includegraphics[scale = 0.4]{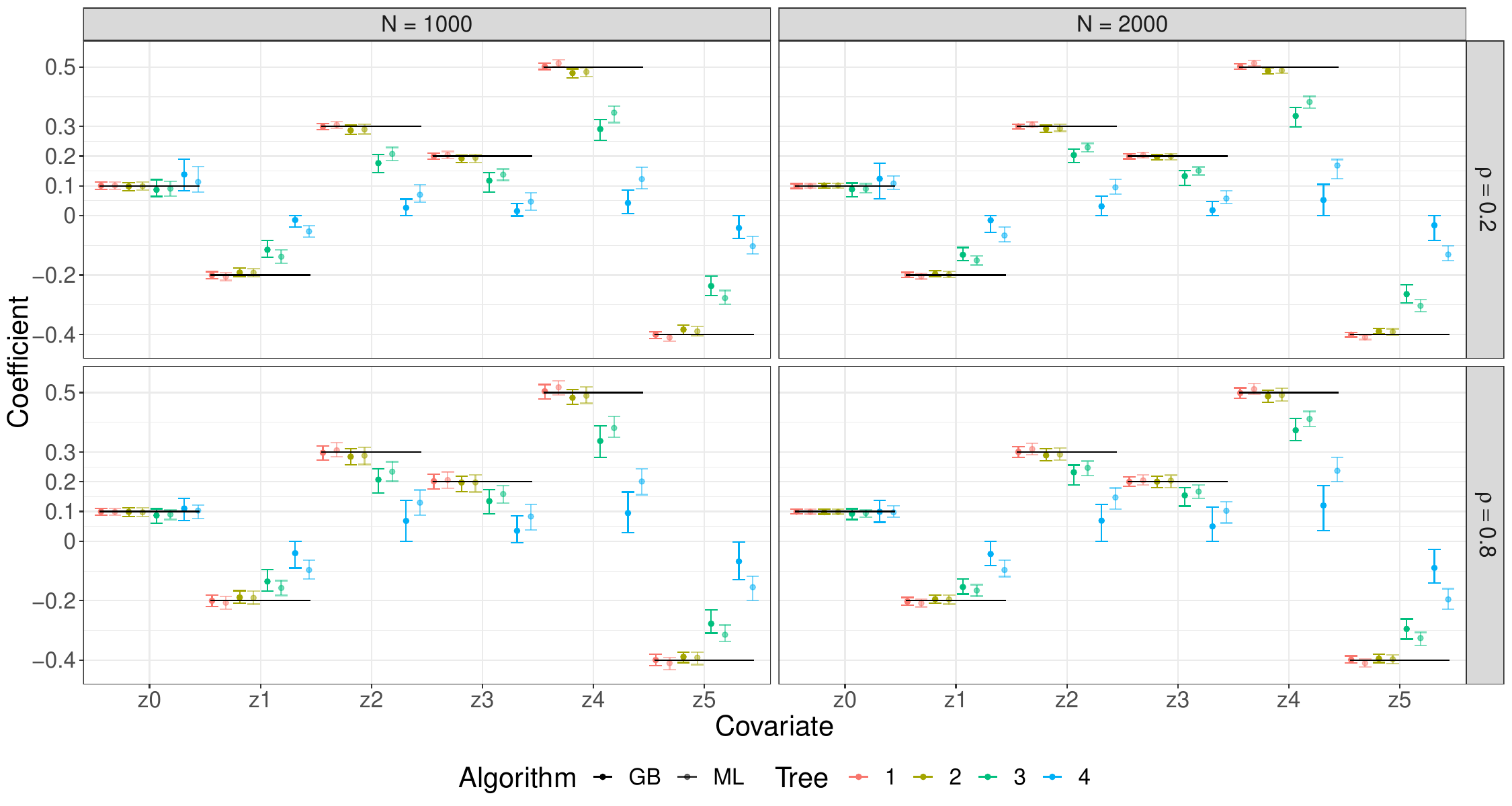}
	\end{center}	
	\caption{Summary of the estimated coefficients based on gradient-boosting (GB, Algorithm \ref{alg: Gradient-boosting}) and (penalized) maximum likelihood (ML, \cite{Vatter2015}). The copula families and the informative covariates have been selected in advance. The bars represent the interquartile range and the points denote the median of the estimated coefficients. The horizontal black lines correspond to the true coefficients. The color represent the tree level while the color transparency stands for the estimation algorithm.}
	\label{fig: GB_vs_ML}
\end{figure}

\begin{table}[h!]
\begin{center}

\begin{tabular}{|c|c|c|} 
\hline
Model & Selected & Specified\\ \hline
low-dimensional & 29 & 6\\
high-dimensional & 106 & 11\\ \hline
\end{tabular}

\caption{Mean computation times (in seconds) averaged over the correlation settings $\rho \in \mng{0.2, 0.8}$ and sample sizes $N\in \mng{1000, 2000}$.}
\label{tab: time_rvine}
\end{center}
\end{table}

\newpage
\pagebreak
\clearpage
\section{Application}
\label{app: Application}

\textcite{Ziel2019} propose the copula based variogram score ($\text{CVS}^{p}$) and energy score (CES) as an extension of existent multivariate scoring rules (see, e.g. \cites{Scheuerer2015b, Gneiting2007}). The copula based multivariate scoring rules allow to evaluate the forecast skill of the multivariate dependence independently of the marginal distributions. For more details on the definition of these scores and further information, we refer to \textcite{Ziel2019}.

\begin{table}[h!]
\begin{center}
\begin{tabular}{c c c c c c} 
\toprule
 & Raw ensemble & ECC & GCA & CVC-G & CVC \\ \hline
$\text{CVS}^{0.5}$ & $0.964^{\,(0.013,\, \underline{0.012})}$ & $0.967^{\,(0.004,\, \underline{0.003})}$ & $0.956^{\,(0.118,\, \underline{0.122})}$ & \textbf{0.951}$^{\,(\underline{0.539})}$ & $0.951^{\,(0.0461)}$ \\ 
CES & $0.177^{\,(0.000,\, \underline{0.000})}$ & $0.178^{\,(0.000,\, \underline{0.000})}$ & $0.175^{\,(0.276,\, \underline{0.255})}$ & $0.175^{\,(\underline{0.278})}$ & \textbf{0.175}$^{\,(0.722)}$ \\
\bottomrule
\end{tabular}
\end{center}
\caption{Copula based verification scores aggregated over all time points in the validation data set. Bold values represent the best value for each score. The score exponents show the $p$-values in favor of the CVC-G (not underlined) and CVC (underlined) over the competing methods according to the Diebold-Mariano test.}
\label{tab: scores}
\end{table}

For assessing the calibration of multivariate forecasts,  multivariate verification rank histograms (see, e.g. \cites{Gneiting2008, Thorarinsdottir2016}) are frequently employed. We use in the following the multivariate ranking suggested by \textcite{Gneiting2008}, where we iteratively calculate the multivariate ranks of GCA, CVC-G and CVC for ensemble size $m=50$ of all 10000 samples. The closer the distribution of the multivariate ranks to the discrete uniform distribution on $\mng{1,\ldots, m+1}$ is, the better the calibration, where any deviation from uniformity indicates a lack of calibration. The deviation from uniformity can be further assessed  by the reliability index \parencite{Monache2006}, denoted by $\triangle$, where a lower value represents a better calibration. 

\begin{figure}[h!]
	\begin{center}
		\includegraphics[scale = 0.45]{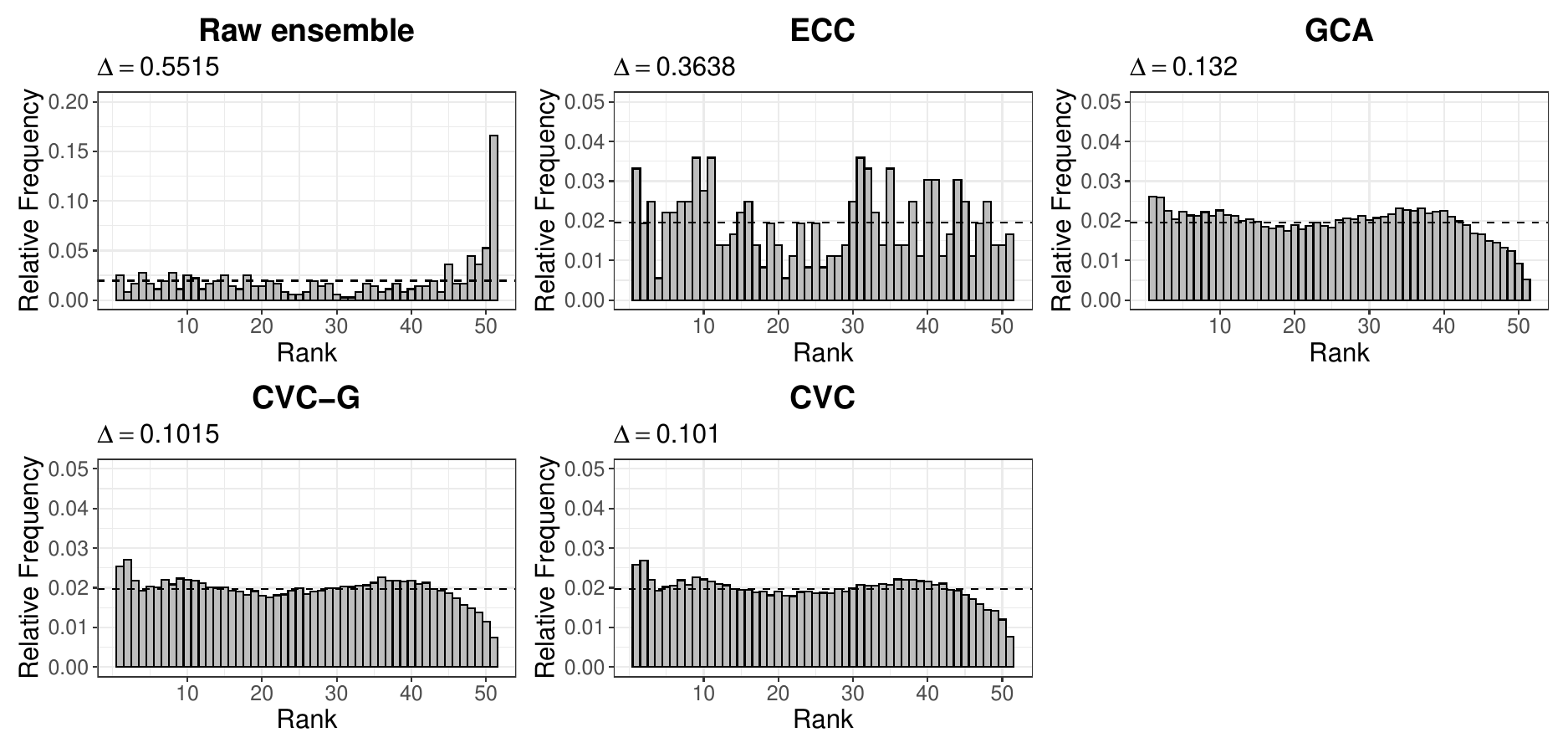}
	\end{center}	
	\caption{Multivariate verification rank histograms for the validation data set.}
	\label{fig: mvrh}
\end{figure}

\flushbottom



\newpage
\addcontentsline{toc}{section}{References}
\thispagestyle{plain}
\clearpage

\printbibliography

@Book{Vannitsem2018,
  author    = {Vannitsem, S. and Wilks, Daniel and Messner, J. W.},
  publisher = {Elsevier},
  title     = {Statistical Postprocessing of Ensemble Forecasts},
  year      = {2018},
  doi       = {10.1016/c2016-0-03244-8},
}

@Manual{RCT2020,
  title        = {R: A Language and Environment for Statistical Computing},
  address      = {Vienna, Austria},
  author       = {{R Core Team}},
  organization = {R Foundation for Statistical Computing},
  year         = {2020},
  url          = {https://www.R-project.org/},
}

@Article{Gneiting2005,
  author    = {Tilmann Gneiting and Adrian E. Raftery and Anton H. Westveld and Tom Goldman},
  journal   = {Monthly Weather Review},
  title     = {Calibrated Probabilistic Forecasting Using Ensemble Model Output Statistics and Minimum {CRPS} Estimation},
  year      = {2005},
  number    = {5},
  pages     = {1098--1118},
  volume    = {133},
  doi       = {10.1175/mwr2904.1},
  publisher = {American Meteorological Society},
}

@Article{Monache2006,
  author    = {Luca Delle Monache and Joshua P. Hacker and Yongmei Zhou and Xingxiu Deng and Roland B. Stull},
  journal   = {Journal of Geophysical Research},
  title     = {Probabilistic aspects of meteorological and ozone regional ensemble forecasts},
  year      = {2006},
  number    = {D24},
  volume    = {111},
  doi       = {10.1029/2005jd006917},
  publisher = {American Geophysical Union ({AGU})},
}

@Article{Gneiting2007,
  author    = {Tilmann Gneiting and Adrian E Raftery},
  journal   = {Journal of the American Statistical Association},
  title     = {Strictly Proper Scoring Rules, Prediction, and Estimation},
  year      = {2007},
  number    = {477},
  pages     = {359--378},
  volume    = {102},
  doi       = {10.1198/016214506000001437},
  publisher = {Informa {UK} Limited},
}

@Article{Gneiting2008,
  author    = {Tilmann Gneiting and Larissa I. Stanberry and Eric P. Grimit and Leonhard Held and Nicholas A. Johnson},
  journal   = {{TEST}},
  title     = {Assessing probabilistic forecasts of multivariate quantities, with an application to ensemble predictions of surface winds},
  year      = {2008},
  number    = {2},
  pages     = {211--235},
  volume    = {17},
  doi       = {10.1007/s11749-008-0114-x},
  publisher = {Springer Science and Business Media {LLC}},
}

@Article{Diebold1995,
  author    = {Francis X. Diebold and Roberto S. Mariano},
  journal   = {Journal of Business {\&} Economic Statistics},
  title     = {Comparing Predictive Accuracy},
  year      = {1995},
  number    = {3},
  pages     = {253--263},
  volume    = {13},
  doi       = {10.1080/07350015.1995.10524599},
  publisher = {Informa {UK} Limited},
}

@Article{Matheson1976,
  author    = {James E. Matheson and Robert L. Winkler},
  journal   = {Management Science},
  title     = {Scoring Rules for Continuous Probability Distributions},
  year      = {1976},
  number    = {10},
  pages     = {1087--1096},
  volume    = {22},
  doi       = {10.1287/mnsc.22.10.1087},
  publisher = {Institute for Operations Research and the Management Sciences ({INFORMS})},
}

@Book{Czado2019,
  author    = {Claudia Czado},
  publisher = {Springer International Publishing},
  title     = {Analyzing Dependent Data with Vine Copulas},
  year      = {2019},
  doi       = {10.1007/978-3-030-13785-4},
}

@Article{Aas2009,
  author = {Kjersti Aas and Claudia Czado and Arnoldo Frigessi and Henrik Bakken},
  title  = {Pair-copula constructions of multiple dependence},
  year   = {2009},
  issn   = {0167-6687},
  pages  = {182-198},
  volume = {44},
  doi    = {10.1016/j.insmatheco.2007.02.001},
}

@Article{Bedford2001,
  author    = {T. Bedford and R. M. Cooke},
  journal   = {Annals of Mathematics and Artificial Intelligence},
  title     = {Probability Density Decomposition for Conditionally Dependent Random Variables Modeled by Vines},
  year      = {2001},
  number    = {1/4},
  pages     = {245--268},
  volume    = {32},
  doi       = {10.1023/a:1016725902970},
  publisher = {Springer Science and Business Media {LLC}},
}

@Article{Bedford2002,
  author    = {T. Bedford and R. M. Cooke},
  journal   = {The Annals of Statistics},
  title     = {Vines{\textemdash}A New Graphical Model for Dependent Random Variables},
  year      = {2002},
  number    = {4},
  pages     = {1031--1068},
  volume    = {30},
  doi       = {10.1214/aos/1031689016},
  publisher = {Institute of Mathematical Statistics},
}

@Article{Dissmann2013,
  author    = {J. Di{\ss}mann and E.C. Brechmann and C. Czado and D. Kurowicka},
  journal   = {Computational Statistics {\&} Data Analysis},
  title     = {Selecting and estimating regular vine copulae and application to financial returns},
  year      = {2013},
  pages     = {52--69},
  volume    = {59},
  doi       = {10.1016/j.csda.2012.08.010},
  publisher = {Elsevier {BV}},
}

@Article{Stoeber2013,
  author    = {Jakob St\"ober and Harry Joe and Claudia Czado},
  journal   = {Journal of Multivariate Analysis},
  title     = {Simplified pair copula constructions{\textemdash}Limitations and extensions},
  year      = {2013},
  pages     = {101--118},
  volume    = {119},
  doi       = {10.1016/j.jmva.2013.04.014},
  publisher = {Elsevier {BV}},
}

@InCollection{Joe1996a,
  author    = {Harry Joe},
  booktitle = {Institute of Mathematical Statistics Lecture Notes - Monograph Series},
  publisher = {Institute of Mathematical Statistics},
  title     = {Families of $m$-variate distributions with given margins and $m(m-1)/2$ bivariate dependence parameters},
  year      = {1996},
  pages     = {120--141},
  doi       = {10.1214/lnms/1215452614},
}

@MastersThesis{Brechmann2010,
  author = {Eike Christian Brechmann},
  school = {Technical University of Munich},
  title  = {Truncated and simplified reulgar vines and their applications},
  year   = {2010},
  type   = {mathesis},
  url    = {https://mediatum.ub.tum.de/doc/1079285/1079285.pdf},
}

@Article{Sklar1959,
  author  = {Sklar, Abe},
  journal = {Publications de L'Institut de Statistique de L'Universit\'e de Paris},
  title   = {Fonctions de R\'epartition \`a Dimensions et Leurs Marges},
  year    = {1959},
  pages   = {229--231},
  volume  = {8},
}

@InCollection{Akaike1998,
  author    = {Hirotogu Akaike},
  booktitle = {Springer Series in Statistics},
  publisher = {Springer New York},
  title     = {Information Theory and an Extension of the Maximum Likelihood Principle},
  year      = {1998},
  pages     = {199--213},
  doi       = {10.1007/978-1-4612-1694-0_15},
}

@Article{Moeller2013,
  author    = {Annette M\"oller and Alex Lenkoski and Thordis L. Thorarinsdottir},
  journal   = {Quarterly Journal of the Royal Meteorological Society},
  title     = {Multivariate probabilistic forecasting using ensemble Bayesian model averaging and copulas},
  year      = {2013},
  number    = {673},
  pages     = {982--991},
  volume    = {139},
  doi       = {10.1002/qj.2009},
  publisher = {Wiley},
}

@Article{Schefzik2013,
  author    = {Roman Schefzik and Thordis L. Thorarinsdottir and Tilmann Gneiting},
  journal   = {Statistical Science},
  title     = {Uncertainty Quantification in Complex Simulation Models Using Ensemble Copula Coupling},
  year      = {2013},
  number    = {4},
  volume    = {28},
  doi       = {10.1214/13-sts443},
  publisher = {Institute of Mathematical Statistics},
}

@Article{Lang2020,
  author    = {Moritz N. Lang and Sebastian Lerch and Georg J. Mayr and Thorsten Simon and Reto Stauffer and Achim Zeileis},
  journal   = {Nonlinear Processes in Geophysics},
  title     = {Remember the past: a comparison of time-adaptive training schemes for non-homogeneous regression},
  year      = {2020},
  month     = {feb},
  number    = {1},
  pages     = {23--34},
  volume    = {27},
  doi       = {10.5194/npg-27-23-2020},
  publisher = {Copernicus {GmbH}},
}

@Article{Hastie1986,
  author    = {Trevor Hastie and Robert Tibshirani},
  journal   = {Statistical Science},
  title     = {Generalized Additive Models},
  year      = {1986},
  month     = {aug},
  number    = {3},
  volume    = {1},
  doi       = {10.1214/ss/1177013604},
  publisher = {Institute of Mathematical Statistics},
}

@Misc{ECMWF2021,
  author = {{European Centre for Medium-Range Weather Forecasts (ECMWF)}},
  note   = {Creative Commons Attribution 4.0 International (CC BY 4.0)},
  title  = {Gridded forecast},
  year   = {2021},
  url    = {https://www.ecmwf.int},
}

@Book{Hastie1990,
  author    = {T.J. Hastie and R.J. Tibshirani},
  publisher = {Taylor \& Francis},
  title     = {Generalized Additive Models},
  year      = {1990},
  month     = jun,
}

@PhdThesis{Patton2002,
  author = {Patton, A. J.},
  school = {University of California},
  title  = {Applications of Copula Theory in Financial Econometrics},
  year   = {2002},
}

@Article{Gijbels2011,
  author    = {Irene Gijbels and Noel Veraverbeke and Marel Omelka},
  journal   = {Computational Statistics \& Data Analysis},
  title     = {Conditional copulas, association measures and their applications},
  year      = {2011},
  month     = {may},
  number    = {5},
  pages     = {1919--1932},
  volume    = {55},
  doi       = {10.1016/j.csda.2010.11.010},
  publisher = {Elsevier {BV}},
}

@Article{Acar2010,
  author    = {Elif F. Acar and Radu V. Craiu and Fang Yao},
  journal   = {Biometrics},
  title     = {Dependence Calibration in Conditional Copulas: A Nonparametric Approach},
  year      = {2010},
  month     = {aug},
  number    = {2},
  pages     = {445--453},
  volume    = {67},
  doi       = {10.1111/j.1541-0420.2010.01472.x},
  publisher = {Wiley},
}

@Article{Vatter2015,
  author    = {Thibault Vatter and Val{\'{e}}rie Chavez-Demoulin},
  journal   = {Journal of Multivariate Analysis},
  title     = {Generalized additive models for conditional dependence structures},
  year      = {2015},
  month     = {oct},
  pages     = {147--167},
  volume    = {141},
  doi       = {10.1016/j.jmva.2015.07.003},
  publisher = {Elsevier {BV}},
}

@Article{Vatter2018,
  author    = {Thibault Vatter and Thomas Nagler},
  journal   = {Journal of Computational and Graphical Statistics},
  title     = {Generalized Additive Models for Pair-Copula Constructions},
  year      = {2018},
  month     = {jul},
  number    = {4},
  pages     = {715--727},
  volume    = {27},
  doi       = {10.1080/10618600.2018.1451338},
  publisher = {Informa {UK} Limited},
}

@Article{Panagiotelis2012,
  author    = {Anastasios Panagiotelis and Claudia Czado and Harry Joe},
  journal   = {Journal of the American Statistical Association},
  title     = {Pair Copula Constructions for Multivariate Discrete Data},
  year      = {2012},
  month     = {may},
  number    = {499},
  pages     = {1063--1072},
  volume    = {107},
  doi       = {10.1080/01621459.2012.682850},
  publisher = {Informa {UK} Limited},
}

@Article{Messner2016,
  author  = {Jakob W. Messner and Georg J. Mayr and Achim Zeileis},
  journal = {{The R Journal}},
  title   = {{Heteroscedastic Censored and Truncated Regression with crch}},
  year    = {2016},
  number  = {1},
  pages   = {173--181},
  volume  = {8},
  doi     = {10.32614/RJ-2016-012},
  url     = {https://doi.org/10.32614/RJ-2016-012},
}

@Dataset{Jobst2023e,
  author    = {Jobst, David and M\"oller, Annette and Gro{\ss}, J\"urgen},
  doi       = {10.5281/zenodo.8193645},
  month     = {July},
  publisher = {Zenodo},
  title     = {{Data set for the ensemble postprocessing of 2m surface temperature forecasts in Germany for five different lead times}},
  url       = {https://doi.org/10.5281/zenodo.8193645},
  version   = {0.1.0},
  year      = {2023},
}

@Article{Veraverbeke2011,
  author    = {N\"oel Veraverbeke and Marek Omelka and Ir{\`{e}}ne Gijbels},
  journal   = {Scandinavian Journal of Statistics},
  title     = {Estimation of a Conditional Copula and Association Measures},
  year      = {2011},
  month     = {jul},
  number    = {4},
  pages     = {766--780},
  volume    = {38},
  doi       = {10.1111/j.1467-9469.2011.00744.x},
  publisher = {Wiley},
}

@Article{Craiu2012,
  author    = {V. Radu Craiu and Avideh Sabeti},
  journal   = {Journal of Multivariate Analysis},
  title     = {In mixed company: Bayesian inference for bivariate conditional copula models with discrete and continuous outcomes},
  year      = {2012},
  month     = {sep},
  pages     = {106--120},
  volume    = {110},
  doi       = {10.1016/j.jmva.2012.03.010},
  publisher = {Elsevier {BV}},
}

@Article{Sabeti2014,
  author    = {Avideh Sabeti and Mian Wei and Radu V. Craiu},
  journal   = {Stat},
  title     = {Additive models for conditional copulas},
  year      = {2014},
  month     = {mar},
  number    = {1},
  pages     = {300--312},
  volume    = {3},
  doi       = {10.1002/sta4.64},
  publisher = {Wiley},
}

@Article{Radice2015,
  author    = {Rosalba Radice and Giampiero Marra and Ma{\l}gorzata Wojty{\'{s}}},
  journal   = {Statistics and Computing},
  title     = {Copula regression spline models for binary outcomes},
  year      = {2015},
  month     = {jun},
  number    = {5},
  pages     = {981--995},
  volume    = {26},
  doi       = {10.1007/s11222-015-9581-6},
  publisher = {Springer Science and Business Media {LLC}},
}

@Article{Marra2017,
  author    = {Giampiero Marra and Rosalba Radice},
  journal   = {Computational Statistics {\&} Data Analysis},
  title     = {Bivariate copula additive models for location, scale and shape},
  year      = {2017},
  month     = {aug},
  pages     = {99--113},
  volume    = {112},
  doi       = {10.1016/j.csda.2017.03.004},
  publisher = {Elsevier {BV}},
}

@Article{Hans2022,
  author    = {Nicolai Hans and Nadja Klein and Florian Faschingbauer and Michael Schneider and Andreas Mayr},
  journal   = {Biometrics},
  title     = {Boosting distributional copula regression},
  year      = {2022},
  month     = {oct},
  doi       = {10.1111/biom.13765},
  publisher = {Wiley},
}

@Article{Mayr2018,
  author    = {Andreas Mayr and Benjamin Hofner},
  journal   = {Statistical Modelling},
  title     = {Boosting for statistical modelling-A non-technical introduction},
  year      = {2018},
  month     = {jan},
  number    = {3-4},
  pages     = {365--384},
  volume    = {18},
  doi       = {10.1177/1471082x17748086},
  publisher = {{SAGE} Publications},
}

@Article{Buehlmann2003,
  author    = {Peter B\"uhlmann and Bin Yu},
  journal   = {Journal of the American Statistical Association},
  title     = {Boosting With the $L_2$ Loss},
  year      = {2003},
  month     = {jun},
  number    = {462},
  pages     = {324--339},
  volume    = {98},
  doi       = {10.1198/016214503000125},
  publisher = {Informa {UK} Limited},
}

@Article{Buehlmann2007b,
  author    = {Peter B\"uhlmann and Torsten Hothorn},
  journal   = {Statistical Science},
  title     = {Rejoinder: Boosting Algorithms: Regularization, Prediction and Model Fitting},
  year      = {2007},
  month     = {nov},
  number    = {4},
  volume    = {22},
  doi       = {10.1214/07-STS242REJ},
  publisher = {Institute of Mathematical Statistics},
}

@Article{Friedman2001,
  author    = {Jerome H. Friedman},
  journal   = {The Annals of Statistics},
  title     = {Greedy function approximation: A gradient boosting machine.},
  year      = {2001},
  month     = {oct},
  number    = {5},
  volume    = {29},
  doi       = {10.1214/aos/1013203451},
  publisher = {Institute of Mathematical Statistics},
}

@Article{Stroemer2021,
  author    = {Annika Str\"omer and Christian Staerk and Nadja Klein and Leonie Weinhold and Stephanie Titze and Andreas Mayr},
  journal   = {Statistical Methods in Medical Research},
  title     = {Deselection of base-learners for statistical boosting{\textemdash}with an application to distributional regression},
  year      = {2021},
  month     = {dec},
  number    = {2},
  pages     = {207--224},
  volume    = {31},
  doi       = {10.1177/09622802211051088},
  publisher = {{SAGE} Publications},
}

@Book{Wood2017,
  author    = {Simon N. Wood},
  publisher = {Chapman and Hall/{CRC}},
  title     = {Generalized Additive Models},
  year      = {2017},
  month     = {may},
  doi       = {10.1201/9781315370279},
}

@Article{Nelder1972,
  author    = {J. A. Nelder and R. W. M. Wedderburn},
  journal   = {Journal of the Royal Statistical Society. Series A (General)},
  title     = {Generalized Linear Models},
  year      = {1972},
  number    = {3},
  pages     = {370},
  volume    = {135},
  doi       = {10.2307/2344614},
  publisher = {{JSTOR}},
}

@Article{Hofner2012,
  author    = {Benjamin Hofner and Andreas Mayr and Nikolay Robinzonov and Matthias Schmid},
  journal   = {Computational Statistics},
  title     = {Model-based boosting in R: a hands-on tutorial using the R package mboost},
  year      = {2012},
  month     = {dec},
  number    = {1-2},
  pages     = {3--35},
  volume    = {29},
  doi       = {10.1007/s00180-012-0382-5},
  publisher = {Springer Science and Business Media {LLC}},
}

@Article{Hastie2007,
  author    = {Trevor Hastie},
  journal   = {Statistical Science},
  title     = {Comment: Boosting Algorithms: Regularization, Prediction and Model Fitting},
  year      = {2007},
  month     = {nov},
  number    = {4},
  volume    = {22},
  doi       = {10.1214/07-sts242a},
  publisher = {Institute of Mathematical Statistics},
}

@Article{Hofner2016,
  author    = {Benjamin Hofner and Andreas Mayr and Matthias Schmid},
  journal   = {Journal of Statistical Software},
  title     = {gamboostLSS: An R Package for Model Building and Variable Selection in the GAMLSS Framework},
  year      = {2016},
  number    = {1},
  volume    = {74},
  doi       = {10.18637/jss.v074.i01},
  publisher = {Foundation for Open Access Statistic},
}

@Article{Klein2015,
  author    = {Nadja Klein and Thomas Kneib},
  journal   = {Statistics and Computing},
  title     = {Simultaneous inference in structured additive conditional copula regression models: a unifying Bayesian approach},
  year      = {2015},
  month     = {jun},
  number    = {4},
  pages     = {841--860},
  volume    = {26},
  doi       = {10.1007/s11222-015-9573-6},
  publisher = {Springer Science and Business Media {LLC}},
}

@Article{Lakatos2023,
  author    = {M{\'{a}}ria Lakatos and Sebastian Lerch and Stephan Hemri and S{\'{a}}ndor Baran},
  journal   = {Quarterly Journal of the Royal Meteorological Society},
  title     = {Comparison of multivariate post-processing methods using global {ECMWF} ensemble forecasts},
  year      = {2023},
  month     = {feb},
  number    = {752},
  pages     = {856--877},
  volume    = {149},
  doi       = {10.1002/qj.4436},
  publisher = {Wiley},
}

@Misc{DWD2018,
  author       = {{DWD Climate Data Center (CDC)}},
  title        = {Historische st\"undliche {S}tationsmessungen der {L}ufttemperatur und {L}uftfeuchte f\"ur {D}eutschland, {V}ersion v006},
  organization = {DWD Climate Data Center (CDC)},
  url          = {https://opendata.dwd.de/climate_environment/CDC/observations_germany/climate/hourly/air_temperature/historical/BESCHREIBUNG_obsgermany_climate_hourly_tu_historical_de.pdf},
  year         = {2018},
}

@Article{Hu2016,
  author    = {Yiming Hu and Maurice J. Schmeits and Schalk Jan van Andel and Jan S. Verkade and Min Xu and Dimitri P. Solomatine and Zhongmin Liang},
  journal   = {Journal of Hydrometeorology},
  title     = {A Stratified Sampling Approach for Improved Sampling from a Calibrated Ensemble Forecast Distribution},
  year      = {2016},
  month     = {sep},
  number    = {9},
  pages     = {2405--2417},
  volume    = {17},
  doi       = {10.1175/jhm-d-15-0205.1},
  publisher = {American Meteorological Society},
}

@Article{Pinson2012b,
  author    = {P. Pinson and R. Girard},
  journal   = {Applied Energy},
  title     = {Evaluating the quality of scenarios of short-term wind power generation},
  year      = {2012},
  month     = {aug},
  pages     = {12--20},
  volume    = {96},
  doi       = {10.1016/j.apenergy.2011.11.004},
  publisher = {Elsevier {BV}},
}

@Article{Scheuerer2015b,
  author    = {Michael Scheuerer and Thomas M. Hamill},
  journal   = {Monthly Weather Review},
  title     = {Variogram-Based Proper Scoring Rules for Probabilistic Forecasts of Multivariate Quantities},
  year      = {2015},
  month     = {mar},
  number    = {4},
  pages     = {1321--1334},
  volume    = {143},
  doi       = {10.1175/mwr-d-14-00269.1},
  publisher = {American Meteorological Society},
}

@Article{Thorarinsdottir2016,
  author    = {Thordis L. Thorarinsdottir and Michael Scheuerer and Christopher Heinz},
  journal   = {Journal of Computational and Graphical Statistics},
  title     = {Assessing the Calibration of High-Dimensional Ensemble Forecasts Using Rank Histograms},
  year      = {2016},
  month     = {jan},
  number    = {1},
  pages     = {105--122},
  volume    = {25},
  doi       = {10.1080/10618600.2014.977447},
  publisher = {Informa {UK} Limited},
}

@Article{Ziel2019,
  author        = {Florian Ziel and Kevin Berk},
  title         = {Multivariate Forecasting Evaluation: On Sensitive and Strictly Proper Scoring Rules},
  year          = {2019},
  month         = oct,
  archiveprefix = {arXiv},
  eprint        = {1910.07325},
  primaryclass  = {stat.ME},
}

@Article{Thomas2017,
  author    = {Janek Thomas and Andreas Mayr and Bernd Bischl and Matthias Schmid and Adam Smith and Benjamin Hofner},
  journal   = {Statistics and Computing},
  title     = {Gradient boosting for distributional regression: faster tuning and improved variable selection via noncyclical updates},
  year      = {2017},
  month     = {may},
  number    = {3},
  pages     = {673--687},
  volume    = {28},
  doi       = {10.1007/s11222-017-9754-6},
  publisher = {Springer Science and Business Media {LLC}},
}

@Manual{Papadakis2021,
  title  = {Rfast: A Collection of Efficient and Extremely Fast R Functions},
  author = {Manos Papadakis and Michail Tsagris and Marios Dimitriadis and Stefanos Fafalios and Ioannis Tsamardinos andMatteo Fasiolo and Giorgos Borboudakis and John Burkardt and Changliang Zou and Kleanthi Lakiotaki and Christina Chatzipantsiou.},
  note   = {R package version 2.0.3},
  year   = {2021},
  url    = {https://CRAN.R-project.org/package=Rfast},
}

@Article{Genest2009,
  author    = {Christian Genest and Hans U. Gerber and Marc J. Goovaerts and Roger J.A. Laeven},
  journal   = {Insurance: Mathematics and Economics},
  title     = {Editorial to the special issue on modeling and measurement of multivariate risk in insurance and finance},
  year      = {2009},
  month     = {apr},
  number    = {2},
  pages     = {143--145},
  volume    = {44},
  doi       = {10.1016/j.insmatheco.2008.10.005},
  publisher = {Elsevier {BV}},
}

@Article{Buehlmann2007,
  author    = {B\"uhlmann, Peter and Hothorn, Torsten},
  journal   = {Statistical Science},
  title     = {Boosting Algorithms: Regularization, Prediction and Model Fitting},
  year      = {2007},
  issn      = {0883-4237},
  month     = nov,
  number    = {4},
  volume    = {22},
  doi       = {10.1214/07-sts242},
  publisher = {Institute of Mathematical Statistics},
}

@Book{McNeil2005,
  author    = {McNeil, A.J. and Frey, R. and Embrechts, P.},
  publisher = {Princeton University Press},
  title     = {Quantitative Risk Management},
  year      = {2005},
  address   = {Princeton},
}

@Article{Smith2010,
  author    = {Smith, Michael and Min, Aleksey and Almeida, Carlos and Czado, Claudia},
  journal   = {Journal of the American Statistical Association},
  title     = {Modeling Longitudinal Data Using a Pair-Copula Decomposition of Serial Dependence},
  year      = {2010},
  issn      = {1537-274X},
  month     = dec,
  number    = {492},
  pages     = {1467--1479},
  volume    = {105},
  doi       = {10.1198/jasa.2010.tm09572},
  publisher = {Informa UK Limited},
}

@InBook{NaiRuscone2016,
  author    = {Nai Ruscone, Marta and Osmetti, Silvia Angela},
  pages     = {373--380},
  publisher = {Springer International Publishing},
  title     = {Modelling the Dependence in Multivariate Longitudinal Data by Pair Copula Decomposition},
  year      = {2016},
  isbn      = {9783319429724},
  month     = jul,
  booktitle = {Soft Methods for Data Science},
  doi       = {10.1007/978-3-319-42972-4_46},
  issn      = {2194-5365},
}

@Article{Zhang2022,
  author    = {Zhang, Boyao and Hepp, Tobias and Greven, Sonja and Bergherr, Elisabeth},
  journal   = {Computational Statistics},
  title     = {Adaptive step-length selection in gradient boosting for Gaussian location and scale models},
  year      = {2022},
  issn      = {1613-9658},
  month     = jan,
  number    = {5},
  pages     = {2295--2332},
  volume    = {37},
  doi       = {10.1007/s00180-022-01199-3},
  publisher = {Springer Science and Business Media LLC},
}

@Book{Fahrmeir2021,
  author    = {Fahrmeir, Ludwig and Kneib, Thomas and Lang, Stefan and Marx, Brian D.},
  date      = {2021},
  title     = {Regression: Models, Methods and Applications},
  doi       = {10.1007/978-3-662-63882-8},
  isbn      = {9783662638828},
  publisher = {Springer Berlin Heidelberg},
}

@Article{Jobst2024,
  author     = {Jobst, David and M\"oller, Annette and Gro{\ss}, J\"urgen},
  date       = {2024},
  title      = {Time Series based Ensemble Model Output Statistics for Temperature Forecasts Postprocessing},
  doi        = {10.48550/ARXIV.2402.00555},
  eprint     = {2402.00555},
  eprinttype = {arXiv},
  copyright  = {Creative Commons Attribution 4.0 International},
  year       = {2024},
}

@Article{Sanchez2024,
  author      = {Sanchez, Guillermo Brise\tilde{n}o and Klein, Nadja and Klinkhammer, Hannah and Mayr, Andreas},
  date        = {2024-03-04},
  title       = {Boosting Distributional Copula Regression for Bivariate Binary, Discrete and Mixed Responses},
  doi         = {10.48550/ARXIV.2403.02194},
  eprint      = {2403.02194},
  eprintclass = {stat.ME},
  eprinttype  = {arXiv},
  copyright   = {arXiv.org perpetual, non-exclusive license},
  publisher   = {arXiv},
  year        = {2024},
}

@PhdThesis{Stoeber2013b,
  author   = {St\"ober, Jakob},
  title    = {Regular Vine Copulas with the simplifying assumption, time-variation, and mixed discrete and continuous margins},
  language = {en},
  pages    = {160},
  url      = {https://mediatum.ub.tum.de/1137287},
  school   = {Technische Universit\"at M\"unchen},
  year     = {2013},
}

@Article{Newey1994,
  author       = {Newey, W. K. and West, K. D.},
  date         = {1994-10},
  journaltitle = {The Review of Economic Studies},
  title        = {Automatic Lag Selection in Covariance Matrix Estimation},
  doi          = {10.2307/2297912},
  issn         = {1467-937X},
  number       = {4},
  pages        = {631--653},
  volume       = {61},
  publisher    = {Oxford University Press (OUP)},
}


\end{document}